\documentclass[11pt,a4paper]{article}
\pdfoutput=1
\usepackage{jheppub}

\usepackage{graphicx}
\usepackage[utf8]{inputenc} 

\usepackage{amsmath}
\usepackage{amssymb}
\usepackage{float}
\usepackage{comment}
\usepackage{slashed}
\usepackage[normalem]{ulem}
\usepackage{cancel}

\usepackage{multirow}
\usepackage{rotating}

\usepackage[font=small,justification=raggedright,format=plain]{caption}
\usepackage{subcaption}

\usepackage{amsmath}
\usepackage{amssymb}

\def\vb#1{\vbox to #1 pt{}}
\def\ifmath#1{\relax\ifmmode #1\else $#1$\fi}
\def\half{\ifmath{{\textstyle{\frac{1}{2}}}}}

\allowdisplaybreaks

\def\gsim{\raise0.3ex\hbox{$\;>$\kern-0.75em\raise-1.1ex\hbox{$\sim\;$}}}
\def\lsim{\raise0.3ex\hbox{$\;<$\kern-0.75em\raise-1.1ex\hbox{$\sim\;$}}}
\def\lfv{lepton flavour violation }

\def\SM{$\mathrm{SU(3)_c \otimes SU(2)_L \otimes U(1)_Y}$ }
\def\su2{$\mathrm{SU(2)_L}$}
\newcommand{\sm}{{Standard Model }}

\usepackage{soul}

\definecolor{mightnightblue}{RGB}{25,25,112}

\definecolor{brown}{rgb}{0.59, 0.29, 0.0}

\def\lfv{lepton flavour violation }

\def\vev#1{\left\langle #1\right\rangle}
\def\SM{$\mathrm{SU(3)_c \otimes SU(2)_L \otimes U(1)_Y}$ }
\def\21{$\mathrm{SU(2)_L \otimes U(1)_Y}$}
\def\lfv{lepton flavor violation }

\begin{document}


\title{\boldmath Electroweak Breaking and Higgs Boson Profile in the Simplest Linear Seesaw Model }

%
%

\author[a]{Duarte Fontes,}
\author[a]{Jorge C. Rom\~ao,}
\author[b]{and J.~W.~F. Valle}
\affiliation[a]{Departamento de F\'{\i}sica and CFTP,
Instituto Superior T\'{e}cnico, Universidade de Lisboa, \\
Avenida Rovisco Pais 1, 1049-001 Lisboa, Portugal}
\affiliation[b]{AHEP Group, Institut de F\'{i}sica Corpuscular --
  C.S.I.C./Universitat de Val\`{e}ncia, Parc Cient\'ific de Paterna. C/ Catedr\'atico Jos\'e Beltr\'an, 2 E-46980 Paterna (Valencia) - SPAIN}
\emailAdd{duartefontes@tecnico.ulisboa.pt}
\emailAdd{jorge.romao@tecnico.ulisboa.pt}
\emailAdd{valle@ific.uv.es}
\vspace{0.5cm}
\noindent\abstract{%
We examine the simplest realization of the linear seesaw mechanism within the \sm gauge structure.
Besides the standard scalar doublet, there are two lepton-number-carrying scalars, a nearly inert \su2 doublet and a singlet.
Neutrino masses result from the spontaneous violation of lepton number, implying the existence of a Nambu-Goldstone boson.
Such ``majoron'' would be copiously produced in stars, leading to stringent astrophysical constraints. 
We study the profile of the Higgs bosons in this model, including their effective couplings to the vector bosons and their invisible decay branching ratios.
A consistent electroweak symmetry breaking pattern emerges with a compressed spectrum of scalars in which the ``Standard Model'' Higgs boson can have a sizeable invisible decay into the invisible majorons. 
}

\maketitle
\noindent

\section{Introduction}
\label{Sect:intro}

Non-zero neutrino masses constitute one of the most robust evidences for new physics.
Ever since the discovery~\cite{Kajita:2016cak,McDonald:2016ixn} and confirmation~\cite{Eguchi:2002dm,Ahn:2002up} of neutrino oscillations took place, the efforts to underpin the origin of neutrino mass have been fierce.
Yet the basic dynamical understanding of the smallness of neutrino mass remains as elusive as ever.
We have no clue as to what is the nature of the underlying mechanism and its characteristic energy scale.
A popular approach to neutrino mass generation is the type-I seesaw mechanism, in which neutrinos get mass due to the exchange of heavy singlet mediators.
Following Refs.~\cite{Schechter:1980gr,Chikashige:1980ui,Schechter:1981cv} we assume here that the seesaw mechanism is realized using just the \sm (SM) gauge structure associated to the \SM symmetry.

In its standard high-scale realization, the seesaw mechanism hardly leads to any phenomenological implication besides those associated to the neutrino masses themselves.
However, the seesaw can arise from low-scale physics~\cite{Joshipura:1992hp,Boucenna:2014zba}.
For example, the seesaw mechanism can be realized at low scale in two different pathways, the inverse~\cite{Mohapatra:1986bd,GonzalezGarcia:1988rw} and the linear seesaw~\cite{Akhmedov:1995ip,Akhmedov:1995vm,Malinsky:2005bi}. 
These low-scale seesaw schemes require the addition of a sequential pair of isosinglet leptons, instead of just a single right-handed neutrino added sequentially.

In this work, we examine the simplest variant of the linear seesaw mechanism. 
In contrast to the conventional formulations~\cite{Akhmedov:1995ip,Akhmedov:1995vm,Malinsky:2005bi}, here left-right symmetry is not imposed. In our setup the linear seesaw mechanism is realized in terms of the simplest Standard Model \SM gauge structure, in which lepton number symmetry is ungauged.
Spontaneous breaking of the lepton number symmetry hence implies the existence of a Nambu-Golstone boson, a variant of the so-called majoron~\cite{Chikashige:1980ui,Schechter:1981cv}.
Such minimally extended scalar boson sector contains, in addition
  to the Standard Model Higgs doublet, a second Higgs doublet,
  as well as a complex singlet scalar, both carrying lepton number charges. The singlet is required to ensure consistency with the LEP measurement of the invisible Z decay width. To prevent excessive stellar cooling by majoron emission, their vacuum expectation values (vevs) must obey a stringent astrophysical bound on the vev of the second doublet. This nicely fits the generation of neutrino masses by the linear seesaw mechanism. 
Our ``neutrino-motivated'' singlet extension of the two-doublet Higgs sector also leads to a peculiar benchmark for electroweak (EW) breaking studies at collider experiments.

We perform a numerical study of the Higgs sector taking into account consistency with astrophysical bound, EW precision data as well as perturbative unitarity and vacuum stability.  
These imply that the model nearly realizes the structure of the inert Higgs doublet model~\cite{Deshpande:1977rw,LopezHonorez:2006gr}.
Moreover, it also has an impact on the physics of the 125 GeV \sm Higgs boson discovered at the LHC.
%
Indeed, the profile of the Higgs sector is modified by the mixing of new CP-even states that affect its couplings and the presence of new CP-odd scalars.
For example, it implies the existence of a new invisible Higgs decay channel with majoron emission~\cite{Joshipura:1992hp}.
This has phenomenomenological implications for collider experiments~\cite{Romao:1992zx,Eboli:1994bm,DeCampos:1994fi,Romao:1992dc,deCampos:1995ten,
deCampos:1996bg,Diaz:1998zg,Hirsch:2004rw,Hirsch:2005wd,Bonilla:2015uwa,Bonilla:2015jdf} and has indeed been searched by LEP and LHC collaborations~\cite{Sirunyan:2018owy,Aaboud:2019rtt}.
The new signals can be studied in proton proton collisions such as at the High-Luminosity LHC setup, as well as in the next generation of lepton collider experiments
such as CEPC, FCC-ee, ILC and CLIC~\cite{CEPCStudyGroup:2018ghi, Abada:2019zxq,Bambade:2019fyw,deBlas:2018mhx}.
Moreover, the majoron can, in certain circumstances, play the role of Dark Matter~\cite{berezinsky:1993fm,Lattanzi:2007ux,Bazzocchi:2008fh,Lattanzi:2013uza,Lattanzi:2014mia,Kuo:2018fgw,Heeck:2018lkj,Reig:2019sok}, in addition to having other potential astrophysical and cosmological implications~\cite{Boucenna:2014uma,Lazarides:2018aev}.

The paper is organised as follows. In Sec.~\ref{sec:setup} we describe the basic theory setup of the model, while the main features of the Higgs potential and EW breaking sector are described in Sec.~\ref{sec:higgs-potential}. The theoretical and experimental constraints are described in Sections \ref{sec:theor-constr} and \ref{sec:exper-constr}, respectively. The physical profile of the Higgs boson spectra resulting from our numerical scans are presented in Sec.~\ref{sec:profile-higgs-bosons} while results for the invisible Higgs decay branching ratio are given in Sec.~\ref{sec:invis-higgs-decays}. Finally, a summary is presented in Sec.\ref{sec:Conclusions}.

\section{Basic Theory Setup}
\label{sec:setup}

The Yukawa sector contains, besides the three \sm lepton doublets
\begin{equation}
  \label{eq:3}
  L_i =
  \begin{bmatrix}
    \nu_i\\
    l_i
  \end{bmatrix},
\end{equation}
with lepton number 1, three lepton singlets $\nu_i^c$ with lepton
number $-1$ and three lepton singlets $\psi_i$ with lepton
number $1$.  
The resulting Yukawa Lagrangian is given as
\begin{equation}
  \label{eq:4}
  - \mathcal{L}_{\rm Yuk}= h_{ij} L_i^T C \nu^c_j \Phi 
  + M_{ij} \nu^c_i C \psi_j + f_{ij} L_i^T C  \psi_j \chi_L + \text{h.c.}
\end{equation}
where $h_{ij}$ and $f_{ij}$ are dimensionless Yukawa couplings,
$M_{ij}$ is an arbitrary matrix with dimensions of mass, and $\Phi$
and $\chi_L$ are scalar doublets. 

After symmetry breaking it will give the linear seesaw mass matrix,
\begin{equation}
  \label{eq:5}
  M_\nu =
  \begin{bmatrix}
    0&M_D & M_L\\[+2mm]
    M^T_D & 0& M\\[+2mm]
    M_L^T& M^T &0
  \end{bmatrix},
\end{equation}
where the $\mathbf{3\times 3}$ sub-matrices are given as
\begin{equation}
  \label{eq:6}
  M_D = \frac{1}{\sqrt{2}} v_{\phi} h, \ \mathbf{(3\times 3)},\quad
  M_L = \frac{1}{\sqrt{2}} v_L f, \ \mathbf{(3\times 3)},\quad
  M = M \ \mathbf{(3\times 3)},
\end{equation}
where $v_{\phi}$ and $v_L$ represent the vevs of $\Phi$ and $\chi_L$,
respectively. 
Here, the lepton number is broken by the  $M_L\,\nu S$ term. This leads to the effective light neutrino mass matrix given by
\begin{equation}\label{lin}
M_\nu=M_D(M_L M^{-1})^T+(M_L M^{-1}){M_D}^T.
\end{equation}
This matrix scales linearly with respect to the Dirac Yukawa couplings contained in $M_D$, hence giving name to this seesaw mechanism. 
It is clear that this vanishes as $M_L \to 0$, ensuring that the small neutrino masses are ``protected'' by the lepton number symmetry. 
Notice that, in contrast to the original left-right symmetric formulations~\cite{Akhmedov:1995ip,Akhmedov:1995vm,Malinsky:2005bi}, here we realize the linear seesaw mechanism just in terms of the standard \SM gauge structure, having the existence of the majoron as its characteristic feature.
In the next section we analyse the dynamical origin of the  $M_L\,\nu S$ term from a scalar doublet Higgs vacuum expectation value, whose smallness is required by astrophysics and consistent with minimization of the potential. 

The presence of the heavy TeV-scale neutrinos needed to mediate neutrino mass through the linear seesaw leads to a plethora of phenomenological signatures associated to the heavy neutrinos. These include their signatures such as \lfv processes in high energy collisions~\cite{Das:2012ii,Deppisch:2013cya}, as well as \lfv processes at low energies, such as $\mu\to e\gamma$~\cite{Hirsch:2009mx,Forero:2011pc}. Many other aspects of this theory have also been discussed in similar contexts, such as those associated with violation of unitarity of the lepton mixing matrix~\cite{Valle:1987gv,Antusch:2006vwa,Escrihuela:2015wra,Miranda:2016ptb} and its possible impact upon neutrino oscillation experiments~\cite{Miranda:2016wdr}.
In this paper we focus primarily on the profile of the Higgs sector.

\section{The Higgs Potential}
\label{sec:higgs-potential}

We consider two doublets, $\Phi$, $\chi_L$ and a singlet $\sigma$,
\begin{equation}
  \label{eq:1}
  \Phi=
  \begin{bmatrix}
    \phi^+\\[+2mm]
    \frac{1}{\sqrt{2}} \left( v_{\phi} + R_1 + i\, I_1\right)
  \end{bmatrix},
    ~~~\chi_L=
  \begin{bmatrix}
    \chi_L^+\\[+2mm]
    \frac{1}{\sqrt{2}} \left( v_L + R_2 + i\, I_2\right)
  \end{bmatrix},
  ~~~\sigma =     \frac{1}{\sqrt{2}} \left( v_\sigma + R_3 + i\, I_3\right),
\end{equation}
and choose the following lepton number assignments for the Higgs fields,
\begin{equation}
  \label{eq:2}
  L[\Phi] =0,\quad L[\chi_L] =-2,\quad L[\sigma]=1.
\end{equation}

With these quantum number the most general Higgs potential that we can write that respects all the symmetries is
\begin{align}
  \label{eq:7}
  V_{\rm Higgs}=&-\mu^2\, \Phi^\dagger \Phi + \lambda\, \left(\Phi^\dagger
    \Phi\right)^2 -\mu_L^2\, \chi_L^\dagger \chi_L + \lambda_L\,
  \left(\chi_L^\dagger   \chi_L\right)^2
  - \mu_\sigma^2 \sigma^\dagger \sigma
  + \lambda_\sigma  \left(\sigma^\dagger \sigma\right)^2\nonumber\\[+2mm]
  &
  + \beta_1\, \Phi^\dagger \Phi\, \chi_L^\dagger \chi_L
  + \beta_2\, \Phi^\dagger \Phi\,  \sigma^\dagger \sigma
  + \beta_3\,  \chi_L^\dagger \chi_L\,  \sigma^\dagger \sigma
 +  \beta_5\,  \Phi^\dagger \chi_L\,  \chi_L^\dagger \Phi\nonumber\\[+2mm]
  &
  - \beta_4\, \left(\Phi \chi_L \sigma^2 + \text{h.c.}\right) .
\end{align}
For definiteness we assume all couplings to be real.

\subsection{Minimization Conditions}

First we solve the minimization equations for the mass parameters $\mu,\mu_L,\mu_\sigma$ in the potential. We get
\begin{align}
  \label{eq:8}
  \mu^2=&\frac{\beta_1 v_{\phi} v_L^2+\beta_2 v_{\phi}
    v_\sigma^2-\beta_4 v_L v_\sigma^2+2
   \lambda v_\phi^3+ \beta_5 v_L^2 v_\phi}{2 v_{\phi}},
  \nonumber\\[+2mm]
  \mu_L^2=&\frac{\beta_1 v_{\phi}^2 v_L+\beta_3 v_L
    v_{\sigma}^2-\beta_4 v_{\phi} v_{\sigma}^2+2 
   \lambda_L v_L^3 + \beta_5 v_L v_\phi^2}{2 v_L},
  \nonumber\\[+2mm]
  \mu_\sigma^2=&\frac{\beta_2 v_{\phi}^2 v_{\sigma}+\beta_3 v_L^2 v_{\sigma}-2
    \beta_4 v_{\phi} v_L v_{\sigma}+2 
   \lambda_S v_{\sigma}^3}{2 v_{\sigma}}.
\end{align}
\subsection{Charged Mass Matrix}

Substituting the minimization conditions we obtain the charged scalar mass matrix in the basis $(\phi^+,\chi^+_L)$ as
\begin{equation}
  \label{eq:9}
  M^2_{\rm ch}=
  \begin{bmatrix}
    \frac{\beta_4 v_L v_{\sigma}^2-\beta_5 v_L^2 v_\phi}{2 v}
    & \frac{\beta_5 v_L v_\phi - \beta_4 v_{\sigma}^2}{2} \\[+2mm]
    \frac{\beta_5 v_L v_\phi -\beta_4 v_{\sigma}^2}{2}
    & \frac{\beta_4 v_{\phi} v_{\sigma}^2-\beta_5 v_L v_\phi^2}{2 v_L} 
  \end{bmatrix},
\end{equation}
which we can easily see that has a zero eigenvalue corresponding to the charged Goldstone boson,
\begin{equation}
  \label{eq:10}
  G^+ =\frac{1}{\sqrt{v_{\phi}^2+v_L^2}}
  \begin{bmatrix}
    v_{\phi}\\[+2mm]
    v_L
  \end{bmatrix} =
  \begin{bmatrix}
    \cos\beta\\
    \sin\beta
  \end{bmatrix},
\end{equation}
where we have defined, as usual,
\begin{equation}
  \label{eq:16}
 v=\sqrt{v_{\phi}^2+v_L^2}=246\text{GeV},\quad \tan\beta= \frac{v_L}{v_{\phi}}.
\end{equation}
The physical charged Higgs has a mass given by
\begin{equation}
  \label{eq:11}
  m_{H^+}^2
  = \frac{(\beta_4 v_{\sigma}^2 -\beta_5 v_L v_\phi) \left(v_{\phi}^2+v_L^2\right)}{2 v_{\phi} v_L} =
  \frac{\beta_4 v_{\sigma}^2}{\sin 2\beta} - \frac{1}{2} \beta_5 v^2.
\end{equation}

\subsection{Neutral Scalar Matrix}

The neutral scalar mass matrix is given by
\begin{equation}
  \label{eq:12}
  M^2_{\rm ns} =
  \begin{bmatrix}
    2 \lambda v_{\phi}^2+\frac{\beta_4 v_L v_{\sigma}^2}{2 v_{\phi}} & \beta_1 v_{\phi}
   v_L-\frac{\beta_4 v_{\sigma}^2}{2} + \beta_5 v_L v_\phi & (\beta_2 v_{\phi}-\beta_4
   v_L) v_{\sigma} \\[+2mm] 
 \beta_1 v_{\phi} v_L-\frac{\beta_4 v_{\sigma}^2}{2}+ \beta_5 v_L v_\phi & 2 \lambda_L
   v_L^2+\frac{\beta_4 v_{\phi} v_{\sigma}^2}{2 v_L} & \beta_3 v_L
   v_{\sigma}-\beta_4 v_{\phi} v_{\sigma} \\[+2mm]
 (\beta_2 v-\beta_4 v_L) v_{\sigma} & \beta_3 v_L v_{\sigma}-\beta_4 v_{\phi} v_{\sigma} & 2
   \lambda_\sigma v_{\sigma}^2 
  \end{bmatrix},
\end{equation}
and we can check that has non-zero determinant, so there are three massive CP-even scalars.

\subsection{Neutral Pseudo-Scalar Matrix}

The neutral pseudo-scalar mass matrix is given by
\begin{equation}
\label{eq:13}
  M^2_{\rm nps} =
  \begin{bmatrix}
     \frac{\beta_4 v_L v_{\sigma}^2}{2 v_{\phi}} & -\frac{\beta_4 v_{\sigma}^2}{2} & -\beta_4
   v_L v_{\sigma} \\[+2mm]
 -\frac{\beta_4 v_{\sigma}^2}{2} & \frac{\beta_4 v_{\phi} v_{\sigma}^2}{2 v_L} & \beta_4 v_{\phi} v_{\sigma}
   \\[+2mm]
 -\beta_4 v_L v_{\sigma} & \beta_4 v_{\phi} v_{\sigma} & 2 \beta_4 v_{\phi} v_L 
  \end{bmatrix}.
\end{equation}
It is easy to verify that it has zero determinant and two zero eigenvalues; their eigenvectors are given by
\begin{align}
  \label{eq:10}
  G^0 =&\frac{1}{\sqrt{v_{\phi}^2+v_L^2}}
  \begin{bmatrix}
    v\\[+2mm]
    v_L\\[+2mm]
    0
  \end{bmatrix}
=
  \begin{bmatrix}\displaystyle
    \frac{v_{\phi}}{v}\\[+2mm]
    \displaystyle
    \frac{v_L}{v}\\[+2mm]
    0
  \end{bmatrix}=
\begin{bmatrix}
    \cos\beta\\
    \sin\beta\\
    0
  \end{bmatrix},
  \\
  \label{eq:10a}
  J =& \frac{v_{\phi}^2 v_L}
    {\sqrt{\left(v_{\phi}^2+v_L^2\right) \left(v_{\phi}^2 \left(4
   v_L^2+v_{\sigma}^2\right)+v_L^2 v_{\sigma}^2\right)}}
  \begin{bmatrix}
    \frac{2v_L}{v_{\phi}}\\[+2mm]
    -2\\[+2mm]
    \frac{v_{\sigma} (v_{\phi}^2+v_L^2)}{v_{\phi}^2 v_L}
  \end{bmatrix}
  = \frac{v_{\phi} v_L}{v V^2}
  \begin{bmatrix}
    2 v_L\\
    -2 v_\phi\\
    \frac{v_{\sigma} v^2}{v_\phi v_L}
  \end{bmatrix}
  \nonumber\\
  =&
  \frac{\sin 2\beta}{\sqrt{\tan^2\beta' + \sin^2 2\beta}}
  \begin{bmatrix}
    \sin\beta \\
    - \cos\beta\\
    \frac{\tan\beta'}{\sin 2\beta}
  \end{bmatrix},
\end{align}
where we have defined
\begin{equation}
  \label{eq:17}
  v=\sqrt{v_{\phi}^2+v_L^2},\
  V^2=  {\sqrt{v_{\phi}^2 \left(4
        v_L^2+v_{\sigma}^2\right)+v_L^2 v_{\sigma}^2}},\
  \tan\beta = \frac{v_L}{v_\phi},\
    \tan\beta' = \frac{v_{\sigma}}{v}.
\end{equation}
The remaining pseudo-scalar is defined as $A$, and its squared mass is given by

\begin{equation}
  \label{eq:14}
 M_A^2= \frac{\beta_4 \left(4 v_{\phi}^2 v_L^2+v_{\phi}^2 v_{\sigma}^2+v_L^2
   v_{\sigma}^2\right)}{2 v_{\phi} v_L}= \beta_4 \left[ v^2
  \sin 2\beta  + \frac{v_{\sigma}^2}{\sin2\beta} \right].
\end{equation}

\subsection{Parameters of the Lagrangian}

It is useful to write the parameters of the Lagrangian in terms of the physical masses, vevs and the relevant angles of rotation. We have already shown in Eq.~\ref{eq:8} that the quadratic terms of the potential can be written in terms of the vevs and the other parameters. This has already been done in writing the mass squared matrices.

\subsubsection{Neutral Pseudo Scalar Mass Matrix}

It is convenient to write the rotation matrix that connects the weak eigenstates to the mass eigenstates. We get
\begin{equation}
  \label{eq:24}
  \begin{bmatrix}
    G_0\\
    J\\
    A
  \end{bmatrix}
  =\mathcal{O}^{I}
  \begin{bmatrix}
    I_1\\
    I_2\\
    I_3
  \end{bmatrix},
\end{equation}
with
\begin{equation}
 \label{eq:25}
  \text{diag}(0,0,m_{A}^2) =
 \mathcal{O}^{I} \cdot M_{\rm ns}^2\cdot\mathcal{O}^{I}{}^T.
\end{equation}
The matrix $\mathcal{O}^{I}$ is given by
\begin{equation}
  \label{eq:26}
 \mathcal{O}^{I}=
  \begin{bmatrix}
    \cos\beta&\sin\beta&0\\
    \frac{\sin \beta \sin(2\beta)}{\sqrt{\sin ^2(2\beta)+\tan ^2(\beta')}}
    &-\frac{2 \sin \beta
      \cos^2\beta}{\sqrt{\sin^2(2\beta)+\tan ^2(\beta')}}&
    \frac{\tan (\beta')}{\sqrt{\sin ^2(2\beta)+\tan ^2(\beta')}}\\
    -\frac{\sin \beta \tan (\beta')}{\sqrt{\sin^2(2 \beta)+
        \tan^2(\beta')}}
    &\frac{\cos \beta
      \tan (\beta')}{\sqrt{\sin^2(2 \beta)+\tan^2(\beta')}}
    &\frac{\sin (2 \beta)}{\sqrt{\sin^2(2 \beta)+\tan^2(\beta')}}
  \end{bmatrix}.
\end{equation}
Notice that the only parameters involved are again the vevs
    of the scalar multiplets in the Higgs potential. 
We also can invert Eq.~(\ref{eq:14}) to obtain the parameter $\beta_4$ as follows
\begin{equation}
  \label{eq:15}
  \beta_4= \frac{M_A^2 \sin2\beta}{v^2 \left( \sin^22\beta +
      \tan^2\beta'\right)}.
\end{equation}

\subsubsection{Charged Higgs Mass Matrix}

The physical mass of the charged scalar $m_{H^+}$ and pseudo-scalar mass 
$M_A$ are related through
\begin{equation}
  \label{eq:18}
  \beta_5= \frac{2 M_A^2 \tan^2\beta'}{v^2\left(\tan^2\beta' +
      \sin^22\beta\right)}- \frac{2 m_{H^+}^2}{v^2},
\end{equation}
involving, again, the three vevs of the theory.

\subsubsection{Neutral Scalar Mass Matrix}

The diagonalization of the neutral scalar mass matrix will give us six
relations that can be solved to get the parameters
$\lambda,\lambda_L,\lambda_\sigma$ and $\beta_1,\beta_2,\beta_3$
describing the quartic couplings in terms the physical masses, vevs
and rotation angles. We define the scalar mass eigenstates $h_1$,
$h_2$ and $h_3$ as 
\begin{equation}
  \label{eq:19}
  \begin{bmatrix}
    h_1\\
    h_2\\
    h_3
  \end{bmatrix}
  = \mathcal{O}^{R}
  \begin{bmatrix}
    R_1\\
    R_2\\
    R_3
  \end{bmatrix},
\end{equation}
such that
\begin{equation}
  \label{eq:20}
  \mathcal{O}^{R}  \cdot M_{\rm ns}^2\cdot \mathcal{O}^{R}{}^T =
  \text{diag}(M_{1}^2,M_{2}^2,M_{3}^2),
\end{equation}
where $M_1^2$, $M_2^2$ and $M_3^2$ are the squared masses of $h_1$,
$h_2$ and $h_3$, respectively. The matrix $\mathcal{O}^{R}$ can
be parameterized in terms of the angles $\theta_i$ as  
\begin{equation}
  \label{eq:21}
   \mathcal{O}^{R}= \mathcal{O}^{R}_3 \cdot \mathcal{O}^{R}_2 \cdot
   \mathcal{O}^{R}_1  =
  \begin{bmatrix}
    c_1c_2 & s_1 c_2 & s_2\\
    -c_1s_2s_3-s_1c_3&c_1c_3-s_1s_2s_3&c_2s_3\\
    -c_1s_2c_3+s_1s_3&-c_1s_3-s_1s_2c_3&c_2c_3
  \end{bmatrix},
\end{equation}
where
\begin{equation}
  \label{eq:22}
  \mathcal{O}^{R}_1=
  \begin{bmatrix}
    c_1&s_1&0\\
    -s_1&c_1&0\\
    0&0&1
  \end{bmatrix},\quad 
  \mathcal{O}^{R}_2=
  \begin{bmatrix}
    c_2&0&s_2\\
    0&1&0\\
    -s_2&0&c_2
  \end{bmatrix},\quad 
  \mathcal{O}^{R}_3=
  \begin{bmatrix}
    1&0&0\\
    0&c_3&s_3\\
    0&-s_3&c_3
  \end{bmatrix},
\end{equation}
and $c_i=\cos\alpha_i,s_i=\sin\alpha_i$. Using this parameterization one obtains, from Eq.~(\ref{eq:20}), the following relations for the quartic parameters:

\begin{align}
  \lambda=&-\frac{1}{4 v_{\phi}^3}\left[\vb{14}
  \beta_{4} c_{1}^4 v_{L} v_{\sigma}^2+2 c_{1}^2
    \left(\beta_{4} c_{2}^4 
   s_{1}^2 v_{L} v_{\sigma}^2-c_{2}^2 \left(M_{1}^2
     v_{\phi}-2 \beta_{4} 
     s_{1}^2 s_{2}^2 v_{L} v_{\sigma}^2\right)
\right.\right.\nonumber\\
&\left.\left.
  +s_{2}^2
 \left(\beta_{4} c_{3}^4 
   s_{1}^2 s_{2}^2 v_{L} v_{\sigma}^2+c_{3}^2 \left(2 \beta_{4}
     s_{1}^2 
   s_{2}^2 s_{3}^2 v_{L} v_{\sigma}^2-M_{3}^2
   v_{\phi}\right)+\beta_{4} 
   s_{1}^2 s_{2}^2 s_{3}^4 v_{L} v_{\sigma}^2-M_{2}^2 s_{3}^2
   v_{\phi}\right)\right)
\right.\nonumber\\
&\left.
+s_{1}^2 \left(\beta_{4} c_{3}^4
 s_{1}^2 v_{L} 
   v_{\sigma}^2+c_{3}^2 \left(2 \beta_{4} s_{1}^2 s_{3}^2 v_{L}
     v_{\sigma}^2-2 M_{2}^2 
   v_{\phi}\right)+\beta_{4} s_{1}^2 s_{3}^4 v_{L}
 v_{\sigma}^2-2 M_{3}^2 
 s_{3}^2 v_{\phi}\right)
\right.\nonumber\\
&\left.
-4 c_{1} c_{3} s_{1}
 s_{2} s_{3} 
   v_{\phi} (M_{2}^2-M_{3}^2)\vb{14} \right],
 \label{eq:23a}  \\[+2mm]
  \lambda_L=&-\frac{1}{4 v_{L}^3}\left[\vb{14}
  \beta_{4} c_{1}^4 v_{\phi} v_{\sigma}^2+2
    c_{1}^2 \left(\beta_{4} c_{3}^4 
   s_{1}^2 v_{\phi} v_{\sigma}^2+c_{3}^2 \left(2 \beta_{4}
     s_{1}^2 s_{3}^2 
   v_{\phi} v_{\sigma}^2-M_{2}^2 v_{L}\right)+\beta_{4} s_{1}^2
 s_{3}^4 
 v_{\phi} v_{\sigma}^2
\right.\right.\nonumber\\
&\left.\left.
 -M_{3}^2 s_{3}^2
   v_{L}\right)+s_{1}^2 \left(\beta_{4} 
   c_{2}^4 s_{1}^2 v_{\phi} v_{\sigma}^2-2 c_{2}^2 \left(M_{1}^2
   v_{L}-\beta_{4} s_{1}^2 s_{2}^2 v_{\phi}
   v_{\sigma}^2\right)+s_{2}^2 
   \left(\beta_{4} c_{3}^4 s_{1}^2 s_{2}^2 v_{\phi}
     v_{\sigma}^2
\right.\right.\right.\nonumber\\
&\left.\left.\left.
     +c_{3}^2 \left(2 
   \beta_{4} s_{1}^2 s_{2}^2 s_{3}^2 v_{\phi} v_{\sigma}^2-2 M_{3}^2
   v_{L}\right)+\beta_{4} s_{1}^2 s_{2}^2 s_{3}^4
 v_{\phi} v_{\sigma}^2-2 
 M_{2}^2 s_{3}^2 v_{L}\right)\right)
\right.\nonumber\\
&\left.
+4 c_{1}
c_{3} s_{1} 
   s_{2} s_{3} v_{L} (M_{2}^2-M_{3}^2) \vb{14}\right],\label{eq:23b}
  \\[+2mm]
  \lambda_\sigma=&\frac{1}{2 v_{\sigma}^2}\left[\vb{14}
  c_{2}^2 \left(c_{3}^2
      M_{3}^2+M_{2}^2 s_{3}^2\right)+M_{1}^2 
   s_{2}^2 \right],\label{eq:23c}
  \\[+2mm]
  \beta_1=&-\frac{1}{2 v_{L} v_{\phi}}\left[\vb{14}
  c_{1}^4 \left(2 \beta_{5} v_{L}
      v_{\phi}-\beta_{4} v_{\sigma}^2\right)+2 
   c_{1}^2 \left(c_{3}^4 s_{1}^2 \left(2 \beta_{5} v_{L}
       v_{\phi}-\beta_{4} v_{\sigma}^2\right)
\right.\right.\nonumber\\
&\left.\left.
+2 c_{3}^2 s_{1}^2
 s_{3}^2 \left(2 \beta_{5} 
   v_{L} v_{\phi}-\beta_{4} v_{\sigma}^2\right)+s_{1}^2 s_{3}^4
 \left(2 \beta_{5} 
   v_{L} v_{\phi}-\beta_{4} v_{\sigma}^2\right)+c_{3} s_{2} s_{3}
 (M_{2}^2-M_{3}^2)\right)
\right.\nonumber\\
&\left.
+s_{1}^2 \left(c_{3}^4
   s_{1}^2 \left(2 
   \beta_{5} v_{L} v_{\phi}-\beta_{4} v_{\sigma}^2\right)+2 c_{3}^2
 s_{1}^2 
   s_{3}^2 \left(2 \beta_{5} v_{L} v_{\phi}-\beta_{4}
     v_{\sigma}^2\right)+s_{1}^2 
   s_{3}^4 \left(2 \beta_{5} v_{L} v_{\phi}-\beta_{4}
     v_{\sigma}^2\right)
\right.\right.\nonumber\\
&\left.\left.
+2 c_{3} 
   s_{2} s_{3} (M_{3}^2-M_{2}^2)\right)+2 c_{1} s_{1}
   \left(c_{2}^4 \left(c_{3}^2 M_{2}^2+M_{3}^2
   s_{3}^2\right)-c_{2}^2 \left(c_{3}^4
   M_{1}^2
\right.\right.\right.\nonumber\\
&\left.\left.\left.
   +c_{3}^2 \left(2 
   M_{1}^2 s_{3}^2-2 M_{2}^2
   s_{2}^2\right)+M_{1}^2 s_{3}^4-2 
   M_{3}^2 s_{2}^2 s_{3}^2\right)+s_{2}^2 \left(c_{3}^2
   \left(M_{2}^2 s_{2}^2-M_{3}^2\right)
\right.\right.\right.\nonumber\\
&\left.\left.\left.
   +s_{3}^2
   \left(M_{3}^2 
   s_{2}^2-M_{2}^2\right)\right)\right) \vb{14}\right],\label{eq:23d}
  \\[+2mm]
  \beta_2=&\frac{1}{v_{\phi} v_{\sigma}}\left[\vb{14}
  \beta_{4} c_{1}^2 v_{L} v_{\sigma}+s_{1}
    \left(\beta_{4} c_{2}^2 s_{1} 
   v_{L} v_{\sigma}+\beta_{4} s_{1} s_{2}^2 v_{L} v_{\sigma}+c_{2}
   c_{3} s_{3} 
   (M_{3}^2-M_{2}^2)\right)
\right.\nonumber\\
&\left.
 +c_{1} c_{2} s_{2}
 \left(c_{3}^4 
   M_{1}^2-c_{3}^2 \left(M_{3}^2-2 M_{1}^2
     s_{3}^2\right)+M_{1}^2 
   s_{3}^4-M_{2}^2 s_{3}^2\right)\vb{14}\right], \label{eq:23e}
  \\[+2mm]
  \beta_3=&\frac{1}{v_{L} v_{\sigma}}\left[\vb{14}
  \beta_{4} c_{2}^4 v_{\phi} v_{\sigma}+2 \beta_{4}
    c_{2}^2 s_{2}^2 v_{\phi} 
   v_{\sigma}+\beta_{4} s_{2}^4 v_{\phi} v_{\sigma}+c_{1} c_{2}^3
   c_{3} s_{3} 
   (M_{2}^2-M_{3}^2)
 \right.\nonumber\\
 &\left.
   +c_{2} s_{2} \left(c_{1}
     c_{3} s_{2} 
   s_{3} (M_{2}^2-M_{3}^2)+c_{3}^4 M_{1}^2
   s_{1}-c_{3}^2 
   s_{1} \left(M_{3}^2-2 M_{1}^2
     s_{3}^2\right)
 \right.\right.\nonumber\\
&\left.\left.
   +M_{1}^2 s_{1} 
   s_{3}^4-M_{2}^2 s_{1} s_{3}^2\right) \vb{14}\right].\label{eq:23f}
\end{align}

\section{Theoretical Constraints}
\label{sec:theor-constr}

In this section we study the theoretical constraints that must be
applied to the model parameters in order to ensure consistency of the
electroweak symmetry breaking sector. 

\subsection{Stability Constraints }

In order to look at the stability or bounded from below (BFB)
conditions, we start by considering only the neutral vacuum. Defining
$x$, $y$ and $z$ such that: 
\begin{equation}
  \label{eq:46}
\Phi =  \sqrt{x} e^{i \theta_1},~~  \chi_L = \sqrt{y} e^{i \theta_2},~~
 \sigma = \sqrt{z} e^{i \theta_3},
\end{equation}
we can write the quartic terms of the potential \ref{eq:7} as
\begin{equation}
  \label{eq:47}
V_q= V_0 + V_1,
\end{equation}
where
\begin{equation}
  \label{eq:48}
  V_0 = \lambda x^2 + \lambda_L y^2 + \lambda_\sigma z^2
  + 2 \alpha x z + 2 \beta y z + 2\gamma x y,
\end{equation}
with
\begin{equation}
  \label{eq:49}
  \alpha=\half \beta_2, \quad
  \beta=\half \beta_3,\quad
  \gamma=\half \beta_1,
\end{equation}
and
\begin{equation}
  \label{eq:50}
V_1 = 2 |\beta_4| \sqrt{x} \sqrt{y} z\,  \cos ( \delta),
\end{equation}
where $\delta$ is some combination of phases. For the potential of the
form $V_0$, the conditions for stability have been given in
Ref.\cite{Klimenko:1984qx}. The problem is the extra piece
$V_1$. However, we can always say that 
\begin{equation}
  \label{eq:45}
V_1 > V_1^a = -2 |\beta_4| \sqrt{x} \sqrt{y} z  .
\end{equation}
Now, note that for any positive $x,y$, we always have
\begin{equation}
  \label{eq:51}
- \sqrt{x} \sqrt{y} > - x - y  \, .
\end{equation}
Therefore, we can bound our potential in the following way:
\begin{equation}
  \label{eq:52}
  V_1 > V_1^a > V_1^b = -2 |\beta_4| x z  - 2 |\beta_4| y z  ,
\end{equation}
which can be joined into $V_0$ to give
\begin{equation}
  \label{eq:53}
  V > \hat{V} = \lambda x^2 + \lambda_L y^2 + \lambda_\sigma z^2
  + 2 \alpha' x z + 2 \beta' y z + 2\gamma x y  ,
\end{equation}
with
\begin{equation}
  \label{eq:54}
\alpha'=\alpha  - |\beta_4|,\quad
\beta' = \beta - |\beta_4|  .
\end{equation}
Now, from Ref.\cite{Klimenko:1984qx} we get the conditions for the potential to be BFB as follows,
\begin{align}
  \label{eq:55}
  &\left\{\lambda>0, \lambda_L>0,\lambda_\sigma>0;\alpha'>
  -\sqrt{\lambda_\sigma \lambda}; \beta' > -\sqrt{\lambda_\sigma
    \lambda_L}; \gamma > -\sqrt{\lambda \lambda_L};
  \alpha'\ge -\beta' \sqrt{\lambda/\lambda_L} \right\} \nonumber\\
  &\cup \hskip 3mm
  \left\{\lambda>0, \lambda_L>0,\lambda_\sigma>0;
  \sqrt{\lambda_\sigma \lambda_L} \ge \beta' > -\sqrt{\lambda_\sigma
    \lambda_L};
  -\beta' \sqrt{\lambda/\lambda_L} \ge \alpha' > -\sqrt{\lambda_\sigma
    \lambda };\right.\nonumber\\
  &\left.\hskip 10mm
  \lambda_\sigma \gamma> \alpha' \beta' - \sqrt{\Delta_\alpha
    \Delta_\beta}  
  \right\},
\end{align}
where
\begin{equation}
  \label{eq:56}
  \Delta_\alpha = \alpha'^2- \lambda_\sigma \lambda,\quad
  \Delta_\beta = \beta'^2- \lambda_\sigma \lambda_L \ .
\end{equation}
These conditions are sufficient, although they might be more
restrictive than the necessary and sufficient conditions, because of
the method of bounding the potential we have used.

\subsection{Unitarity Constraints}

In order to discuss the unitarity constraints, we follow the procedure
developed in Ref.\cite{Bento:2017eti}. As explained there, we have to
obtain all the coupled channel matrices for the scattering of two
scalars into two scalars, and bound the highest of their eigenvalues.
Since the electric charge and the hypercharge are conserved
in this high energy scattering, we can separate the states according
to these quantum numbers. For this purpose, and because we are in the
very high-energy limit, it is better to work in the unbroken phase. It
is convenient then to use the following notation for the Higgs fields.
\begin{equation}
  \label{equnit:3}
  \Phi=
  \begin{bmatrix}
    w^+_1\\
    n_1
  \end{bmatrix}\, ,
\
  \Phi^\dagger=
  \begin{bmatrix}
    w^-_1\\
    n^*_1
  \end{bmatrix}^T\, ;
  \quad
  \chi_L=
  \begin{bmatrix}
    w^+_2\\
    n_2
  \end{bmatrix}\,  ,
\quad
  \chi_L^\dagger=
  \begin{bmatrix}
    w^-_2\\
    n^*_2
  \end{bmatrix}^T\, ;
  \quad
    \sigma= s\, ,
\quad
  \sigma^*= s^* \, .
\end{equation}
The relevant two body states are given in the
entries of Table~\ref{tab:unitarity1},
and their complex conjugates.
\begin{table}[htb]
  \centering
  \begin{tabular}{|c|c|l|c|}\hline
    $Q$&$Y$&State&Number of states\\[+1mm]\hline\hline
2 &2 &$S_{\alpha}^{++}=\{w_1^+ w_1^+,w_1^+ w_2^+,w_2^+ w_2^+\}$ &
$3$  \\[+1mm]\hline\hline
1 &2 & $S_{\alpha}^{+}=\{w_1^+ n_1,w_1^+ n_2,w_2^+ n_1,w_2^+ n_2\}$
& $4 $ \\[+1mm]\hline
1 &1 & $T_{\alpha}^{+}= \{w_1^+ s,w_1^+ s^*,w_2^+ s,w_2^+ s^*\}$
&$4$ \\[+1mm]\hline
1 &0 & $U_{\alpha}^{+}=  \{w_1^+ n^*_1,w_1^+ n^*_2,w_2^+ n^*_1,w_2^+ n^*_2\} $
&$4$ \\[+1mm]\hline\hline
0 &2 & $S_{\alpha}^{0}=\{n_1 n_1, n_1n_2,n_2n_2\}$
& $3$ \\[+1mm]\hline
    0 &1 & $T_{\alpha}^{0}=\{n_1 s,n_1 s^*,n_2 s,n_2 s^*\} $
& $4$ \\[+1mm]\hline
0 &0 &$U_{\alpha}^{0}=\{w_1^+ w_1^-,w_1^+ w_2^-,w_2^+ w_1^-,w_2^+w_2^-,$
&$11$  \\[+1mm]
 & &$\phantom{U_{\alpha}^{0}=}\hskip +3mm
       n_1 n^*_1,n_1 n^*_2,n_2 n^*_1,n_2 n^*_2,s^* s, s s, s^* s^* \}$
&  \\[+1mm]\hline
  \end{tabular}
  \caption{List of two body scalar states separated by $(Q,Y)$.}
  \label{tab:unitarity1}
\end{table}
It is important to note that the index
$\alpha$ is a compound index;
it refers to a set of $\{i,j\}$ indices for the two body states.
Also note that in Table~\ref{tab:unitarity1} the
two body states with equal particles have a normalization of
$1/\sqrt{2}$ that we have not written here, but must be included in the
calculation. 

\begin{figure}[htb]
    \centering
    \includegraphics{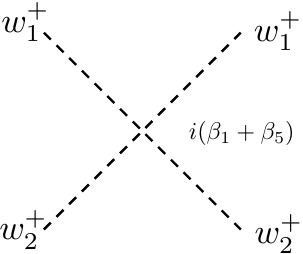}
    \caption{Quartic coupling for $w_1^+ w_2^+ \to w_1^+ w_2^+$.}
    \label{fig:quartic}
  \end{figure}

We will give the full results in the Appendix~\ref{sec:unitarity}, but
let us illustrate with the simplest example, the state
$S_{\alpha}^{++}$. With the notation of Eq.~(\ref{equnit:3}), the
quartic part of the potantial will read
\begin{equation}
  \label{eq:23}
  V_4 = \lambda w_1^+ w_1^- w_1^+ w_1^- + \lambda_L  w_2^+ w_2^-w_2^+w_2^-
  + \beta_1 w_1^+ w_1^- w_2^+ w_2^- + \beta_5 w_1^+ w_2^- w_2^+ w_1^-
  + \cdots
\end{equation}
Now consider the scattering
\begin{equation}
  \label{eq:64}
  w_1^+ w_2^+ \to w_1^+ w_2^+
\end{equation}
This will proceed through the quartic vertex in Fig.~\ref{fig:quartic}.
Therefore the amplitude $\mathcal{M}(w_1^+ w_2^+ \to w_1^+ w_2^+)$
will be given by $\beta_1+\beta_5$. Now consider the scattering
\begin{equation}
  \label{eq:65}
  w_1^+ w_1^+ \to w_1^+ w_1^+
\end{equation}
The Feynman rule for the quartic term would be $4 \lambda$, but
remembering the factor of $1/\sqrt{2}$ for states with identical
particles~\cite{Bento:2017eti} we
would get the amplitude
\begin{equation}
  \label{eq:66}
  \mathcal{M}(\frac{w_1^+ w_1^+}{\sqrt{2}} \to \frac{w_1^+
    w_1^+}{\sqrt{2}}) = 4\lambda \frac{1}{\sqrt{2}}\frac{1}{\sqrt{2}}
  = 2 \lambda
\end{equation}
and similarly for $w_2^+ w_2^+ \to w_2^+ w_2^+$. Therefore for the
coupled channel states in $S_{\alpha}^{++}$ we get,
\begin{equation}
  \label{eq:67}
  16 \pi a_0^{++} = \mathcal{M}^{++} =
  \begin{bmatrix}
    2\lambda & 0 & 0\\
    0& \beta_1 + \beta_5 &0\\
    0& 0& 2\lambda_L
  \end{bmatrix}\, ,
\end{equation}
where $|a_0| < 1/2$, is the partial wave to be bounded, requiring the eigenvalues of Eq.~(\ref{eq:67}) to obey,
\begin{equation}
  \label{eq:68}
  \Lambda_i < 8 \pi\, .
\end{equation}

In the Appendix~\ref{sec:unitarity} we present all the coupled
channel matrices for the sates in Table~\ref{tab:unitarity1} and give
their eigenvalues. Then the limits implied in Eq.~(\ref{eq:68}) were
aplied in the code.

\subsection{Oblique Parameters $S,T,U$}

In order to discuss the effect of the oblique $S,T,U$ parameters, we use the results of Ref.\cite{Grimus:2007if,Branco:2011iw}. To apply their expressions we have to find the matrices $U$ and $V$ that we now define explicitly. For the matrix $U$, we have
\begin{equation}
  \label{eq:27}
  \begin{bmatrix}
  \phi^+\\
  \chi_L^+    
\end{bmatrix}
= U
\begin{bmatrix}
  G^+\\
  H^+
\end{bmatrix},
\end{equation}
which gives
\begin{equation}
  \label{eq:28}
  U=
\begin{bmatrix}
  \cos\beta&-\sin\beta\\
  \sin\beta&\cos\beta
\end{bmatrix}  .
\end{equation}
The matrix $V$ is a $2\times 6$ matrix defined by
\begin{equation}
  \label{eq:29}
  \begin{bmatrix}
    R_1+ i I_1\\
    R_2+i I_2
  \end{bmatrix}=
  V
  \begin{bmatrix}
    G^0\\
    J\\
    A_0\\
    h_1\\
    h_2\\
    h_3
  \end{bmatrix}.
\end{equation}
Using the rotation matrices $\mathcal{O}^{R}$ and $\mathcal{O}^{I}$ we get
\begin{equation}
  \label{eq:30}
  V=
  \begin{bmatrix}
    i\,\mathcal{O}^{I}_{ 11}, i\,\mathcal{O}^{I}_{ 21}, i\,\mathcal{O}^{I}_{ 31}, \mathcal{O}^{R}_{11}, \mathcal{O}^{R}_{21}, \mathcal{O}^{R}_{31}\\
    i\,\mathcal{O}^{I}_{ 12}, i\,\mathcal{O}^{I}_{ 22}, i\,\mathcal{O}^{I}_{ 32}, \mathcal{O}^{R}_{12}, \mathcal{O}^{R}_{22}, \mathcal{O}^{R}_{32}
  \end{bmatrix}.
\end{equation}
To apply the expressions for $S,T,U$, we need the following matrices:
\begin{equation}
  \label{eq:31}
  U^\dagger U =
  \begin{bmatrix}
    1&0\\
    0&1
  \end{bmatrix},
\end{equation}
\begin{equation}
  \label{eq:33}
  \Im(V^\dagger V)=
  \begin{bmatrix}
    0_{3\times 3} & A_{3\times 3}\\
    -A_{3\times 3}& 0_{3\times 3}
  \end{bmatrix},
\end{equation}
where
\begin{equation}
  \label{eq:32}
  A_{3\times 3}=
  \begin{bmatrix}
 -\mathcal{O}^{I}_{11}  \mathcal{O}^{R}_{11} - \mathcal{O}^{I}_{12}  \mathcal{O}^{R}_{12} \ \
& -\mathcal{O}^{I}_{11}  \mathcal{O}^{R}_{21} - \mathcal{O}^{I}_{12}  \mathcal{O}^{R}_{22} \ \
& -\mathcal{O}^{I}_{11}  \mathcal{O}^{R}_{31} - \mathcal{O}^{I}_{12}  \mathcal{O}^{R}_{32} \\
-\mathcal{O}^{I}_{21}  \mathcal{O}^{R}_{11} - \mathcal{O}^{I}_{22}  \mathcal{O}^{R}_{12} \ \
 & -\mathcal{O}^{I}_{21}  \mathcal{O}^{R}_{21} - \mathcal{O}^{I}_{22}  \mathcal{O}^{R}_{22} \ \
 & -\mathcal{O}^{I}_{21}  \mathcal{O}^{R}_{31}  - \mathcal{O}^{I}_{22}  \mathcal{O}^{R}_{32}\\  
 -\mathcal{O}^{I}_{31}  \mathcal{O}^{R}_{11} - \mathcal{O}^{I}_{32}  \mathcal{O}^{R}_{12} \ \
  & -\mathcal{O}^{I}_{31}  \mathcal{O}^{R}_{21} - \mathcal{O}^{I}_{32}  \mathcal{O}^{R}_{22} \ \
  & -\mathcal{O}^{I}_{31}  \mathcal{O}^{R}_{31} - \mathcal{O}^{I}_{32}  \mathcal{O}^{R}_{32}
  \end{bmatrix},
\end{equation}
\begin{equation}
  \label{eq:34}
  U^\dagger V=
  \begin{bmatrix} 
    i& 0& 0& \mathcal{O}^{R}_{11} c_\beta + \mathcal{O}^{R}_{12}
    s_\beta& \mathcal{O}^{R}_{21} c_\beta + \mathcal{O}^{R}_{22}
    s_\beta&  
    \mathcal{O}^{R}_{31} c_\beta + \mathcal{O}^{R}_{32} s_\beta\\
  0& \frac{- i\,  \sin2\beta}{\sqrt{\sin^2(2\beta) +
      \tan\beta'^2}}&  
  \frac{i  \tan\beta'}{\sqrt{\sin^2(2\beta) + \tan\beta'^2}}
  & \mathcal{O}^{R}_{12} c_\beta - \mathcal{O}^{R}_{11} s_\beta& 
  \mathcal{O}^{R}_{22} c_\beta - \mathcal{O}^{R}_{21} s_\beta&
  \mathcal{O}^{R}_{32} c_\beta - \mathcal{O}^{R}_{31} s_\beta 
  \end{bmatrix},
\end{equation}
with $c_\beta=\cos\beta,s_\beta=\sin_\beta$. We also need the diagonal elements
of $V^\dagger V$:
\begin{equation}
  \label{eq:35}
  \text{Diag}(V^\dagger V)=
  \begin{bmatrix}
  1, \frac{\sin2\beta^2}{(\sin^22\beta + \tan\beta'^2)},
  \frac{\tan\beta'^2}{(\sin^22\beta + \tan\beta'^2)}, 
  \mathcal{O}^{R}_{11}{}^2 + \mathcal{O}^{R}_{12}{}^2,
  \mathcal{O}^{R}_{21}{}^2 + \mathcal{O}^{R}_{22}{}^2,  
 \mathcal{O}^{R}_{31}{}^2 + \mathcal{O}^{R}_{32}{}^2   
  \end{bmatrix}.
\end{equation}
We have implemented a numerical code to take all of the above constraints into account.

\section{Experimental Constraints}
\label{sec:exper-constr}

In this section we study the constraints that must be applied to the scalar potential parameters and which follow from various experimental considerations.

\subsection{Astrophysics Constraints}
\label{sec:astr-constr}

Spontaneous breaking of a global symmetry such as lepton number leads to the existence of a Nambu-Goldstone boson, dubbed ``majoron''. 
This would be copiously produced in stars, leading to new mechanisms of stellar cooling.
If the majoron is strictly massless (or lighter than typical stellar temperatures), one has an upper bound for the majoron-electron coupling~\cite{Choi:1989hi,Valle:1990pka} 
\begin{equation} 
  \label{eq:61}
  |g_{Jee}| \lsim 10^{-13} .
\end{equation}
where
\begin{equation}
  \label{eq:62}
|g_{Jee}|=|\vev{ J|\phi}|\frac{m_{e}}{v_{\phi}}  ,
\end{equation}
with $\vev{ J|\phi}$ denoting the majoron projection into the \sm doublet. This can be obtained in a model-independent way by using Noether's theorem~\cite{Schechter:1981cv} and checked explicitly from the form of the pseudo-scalar mass matrix, Eq.~(\ref{eq:10a}). In our case, it leads to the constraint
\begin{equation}
  \label{eq:60}
  |\vev{ J|\phi}|=
\frac{2 v_{\phi} v_L^2}{\sqrt{(v_{\phi}^2+v_L^2)( v_{\phi}^2 (4 v_L^2 +
  v_{\sigma}^2) + v_L^2 v_{\sigma}^2}}
  \lesssim10^{-7}.  
\end{equation}

Note, however, that the majoron can, in certain circumstances, acquire a nonzero mass as a result of interactions explicitly breaking the global lepton number symmetry.
These could arise, say, from quantum gravity effects.
Unfortunately, we have no way of providing a reliable estimate of their magnitude.
If the majoron is massive, it may play the role of warm~\cite{berezinsky:1993fm,Lattanzi:2007ux,Bazzocchi:2008fh,Lattanzi:2013uza,Lattanzi:2014mia,Kuo:2018fgw} or cold
Dark Matter~\cite{Heeck:2018lkj,Reig:2019sok}, in addition to having other potential astrophysical and cosmological implications~\cite{Boucenna:2014uma,Lazarides:2018aev}.
If the majoron mass exceeds the characteristic temperatures of stellar environments, then the bound in  Eq.~(\ref{eq:60}) need not apply. While the missing energy signature associated to the light majoron would remain, there would be important changes in the phenomenological analysis of the scalar sector. 
In what follows, we stick to the validity of Eq.~(\ref{eq:60}), which amounts to having a (nearly) massless majoron.

\subsection{LHC Constraints}

We have to enforce the LHC constraints on the 125 GeV scalar Higgs boson. 
These are given in terms of the so-called signal strength parameters,
\begin{equation}
  \label{eq:16}
  \mu_f =
\frac{\sigma^\textrm{NP}(pp \to h)}{\sigma^\textrm{SM}(pp \to h)}\,
\frac{BR^\textrm{NP}(h \to f)}{BR^\textrm{SM}(h \to f)} ,
\end{equation}
where $\sigma(pp \to h)$ is the cross section for Higgs production, and $BR(h \to f)$ is the branching ratio into the Standard Model final state $f$, with the labels NP and SM denoting New Physics and Standard Model, respectively.  These can be compared with those given by the experimental collaborations. For the 8 TeV data, the signal strengths from a combined ATLAS and CMS analysis~\cite{TheATLASandCMSCollaborations:2015bln} 
are shown in Table~\ref{tab:1}.
\begin{table}[h]
\centering
\begin{tabular}{|ccccccc|}
\hline
channel & & ATLAS  & & CMS & & ATLAS+CMS \\
\hline
$\mu_{\gamma\gamma}$  & &
$1.15^{+0.27}_{-0.25}$  & &
$1.12 ^{+0.25}_{-0.23}$ & &
$1.16^{+0.20}_{-0.18}$
\\*[2mm]
$\mu_{WW}$  & &
$1.23^{+0.23}_{-0.21}$ & &
$0.91^{+0.24}_{-0.21}$  & &
$1.11^{+0.18}_{-0.17}$
\\*[2mm]
$\mu_{ZZ}$  & &
$1.51^{+0.39}_{-0.34}$ & &
$1.05^{+0.32}_{-0.27}$ & &
$1.31^{+0.27}_{-0.24}$
\\*[2mm]
$\mu_{\tau\tau}$  & &
$1.41^{+0.40}_{-0.35}$ & &
$0.89^{+0.31}_{-0.28}$ & &
$1.12^{+0.25}_{-0.23}$
\\*[2mm]
\hline
\end{tabular}
\caption{\label{tab:1} Combined ATLAS and CMS results for the 8 TeV data, Ref.~\cite{TheATLASandCMSCollaborations:2015bln}.} 
\end{table}
For the 13 TeV of Run-2, the data is separated by production process. We took the recent results from ATLAS~\cite{ATLAS-CONF-2018-031} shown in Table~\ref{tab:2}.
\begin{table}[h]
\centering
\begin{tabular}{|c|c|c|c|c|}
\hline
  Decay  & \multicolumn{4}{|c|}{Production Process}  \\
  Mode &  ggF & VBF  & ZH & ttH  \\*[2mm]\hline
  \vb{14} $H\to WW$ &$1.06^{+0.25}_{-0.23}$
       & $1.24^{+0.42}_{-0.37}$ &$0.81^{+0.57}_{-0.49}$
 &$1.27^{+0.44}_{-0.40}$  \\*[3mm]
  $H\to ZZ$ &$1.13^{+0.13}_{-0.13}$ &$1.32^{+0.36}_{-0.31}$
              &$0.86^{+0.58}_{-0.50}$ &$1.36^{+0.38}_{-0.34}$\\*[2mm]
  $H\to \tau\tau$ &$0.98^{+0.33}_{-0.27}$  &$1.15^{+0.48}_{-0.40}$
  &$0.75^{+0.56}_{-0.47}$&$1.18^{+0.50}_{-0.42}$\\*[3mm]
  $H\to \gamma\gamma$ &$1.01^{+0.21}_{-0.19}$  &$1.18^{+0.38}_{-0.33}$
  &$0.76^{+0.54}_{-0.46}$ &$1.21^{+0.40}_{-0.35}$\\*[3mm]
  $H\to bb$ &$0.92^{+0.43}_{-0.33}$
       &$1.07^{+0.57}_{-0.45}$ &$0.70^{+0.57}_{-0.47}$  &$1.10^{+0.59}_{-0.47}$
  \\\hline
\end{tabular}
\caption{\label{tab:2} ATLAS
 results for the 13 TeV
  data, Ref.~\cite{ATLAS-CONF-2018-031}.} 
\end{table}
Finally, we have also enforced the LHC constraints in the other neutral and charged Higgs. This was done using the \texttt{HiggBounds-4} package~\cite{Bechtle:2013wla}. 

In practice, in order to optimize our scans, we started by imposing just the simple requirement that the coupling of the 125 GeV Higgs
scalar boson with the vector bosons, $k_V(h_1)$, lies in the range 
\begin{equation}
  \label{eq:57}
  k_V^2(h_1) \in [0.8,1].
\end{equation}
Applying this first restriction is useful in order to optimize the size of our data points sample. 
Notice that our model has the same structure of the Higgs coupling to the vector bosons as any two Higgs doublet model. This means we can,
for instance, use the results of Ref.~\cite{Fontes:2014xva} to obtain 
\begin{equation}
  \label{eq:58}
  k_V(h_i)=\mathcal{O}^{R}_{i1} c_\beta + \mathcal{O}^{R}_{i2} s_\beta.
\end{equation}
As in the case of any multi-doublet Higgs model, NHDM~\cite{Bento:2017eti}, the couplings of the
CP even Higgs bosons to the vector bosons obey a sum rule,
\begin{align}
  \label{eq:3a}
  \sum_{i=1}^3 |k_V(h_i)|^2 =&  \sum_{i=1}^3 \left[  \mathcal{O}^{R}_{i1} c_\beta +  \mathcal{O}^{R}_{i2} s_\beta \right]^2  =1 
\end{align}
where we have used the properties of the orthogonal matrix $\mathcal{O}^{R}$.
We have started by enforcing Eq.~(\ref{eq:57}) in our scans and
then, to this optimized data set, we applied the constraints on
Eq.~(\ref{eq:16}) using the results from 
Table~\ref{tab:1} and Table~\ref{tab:2}.

\section{The profile of the Higgs bosons}
\label{sec:profile-higgs-bosons}

\subsection{General numerical scan}

\begin{figure}[htb]
  \centering
  \includegraphics[width=0.5\textwidth]{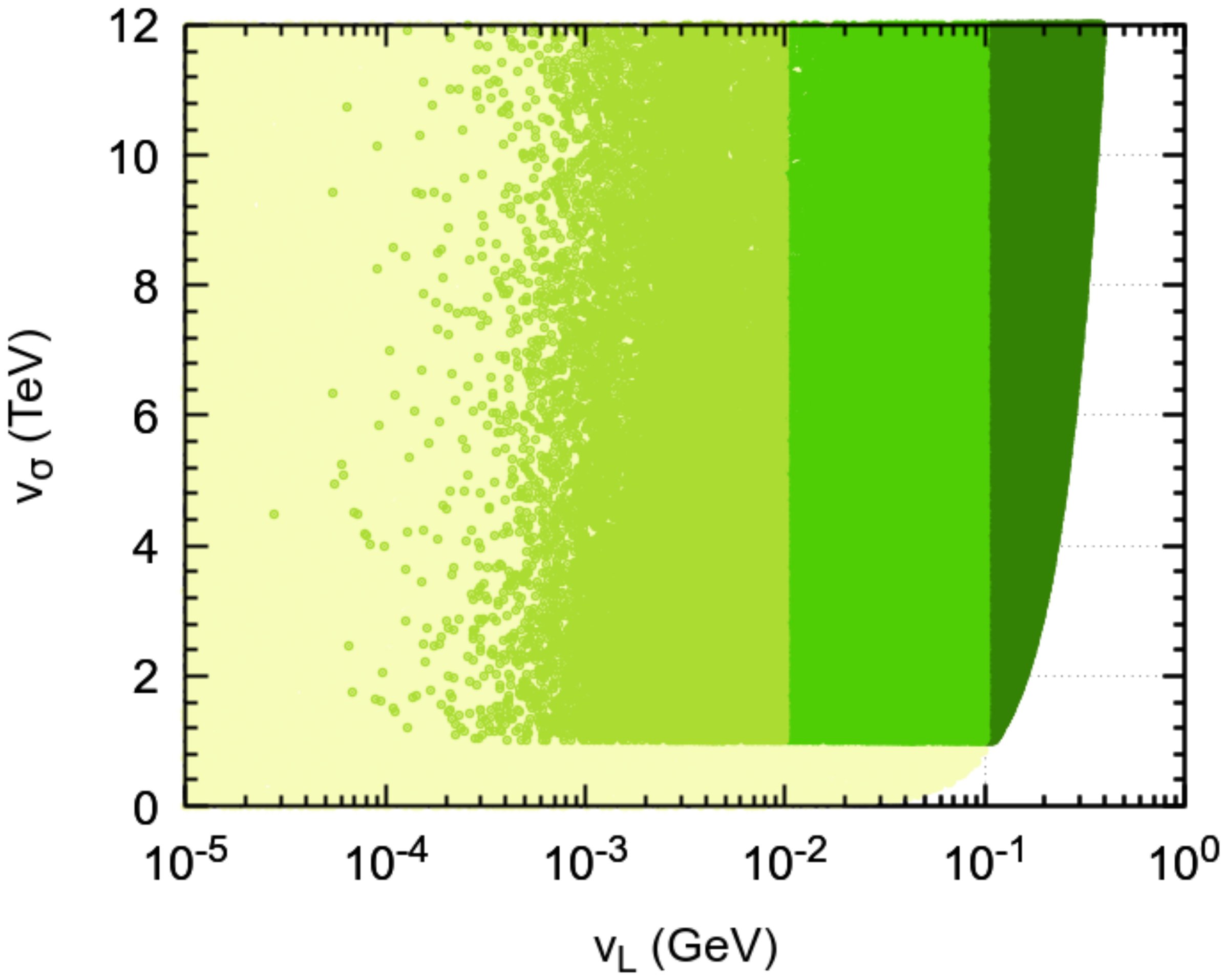}
  \caption{$v_\sigma$ as a function of $v_L$. The yellow region is the set of values of $(v_\sigma,v_L)$ that satisfy Eq.~(\ref{eq:60}) without any other constraint. The other points, in shades of green, satisfy all the other theoretical and experimental constraints. For these, the color code is as follows: the points in dark green have $v_L> 0.1$ GeV, those in green have $v_L\in[0.01,01]$GeV and those in yellow green have $v_L < 0.01$ GeV. In our scans a cut $v_\sigma > 1$ TeV, was imposed. See text for further details.}
  \label{fig:stellar}
\end{figure}

In this section we present a study of the impact of the previous constraints on the parameter space of the scalar potential in the linear seesaw model. In all plots we have imposed the theoretical constraints plus the LHC constraints on $k_V$. We fix the $h_1$ mass to 125 GeV and vary other the model parameters in the following way:\footnote{For technical reasons one can not put $m_J=0$, since there appear logarithms of mass ratios in the evaluation of the S,T,U parameters. However, the limit $m_J\to 0$ is well behaved. To avoid numerical instabilities we took $m_J= 1$ eV, keeping the stellar cooling argument.}
\begin{align}
  \label{eq:59}
  M_{1}=& 125 \, \text{GeV},\quad M_{2}, M_{3},M_{A},M_{H^+} \in
  [125, 800] \, \text{GeV},\quad \alpha_1,\alpha_2,\alpha_3 \in
  [-\frac{\pi}{2},\frac{\pi}{2}]
  \nonumber\\[+2mm]
  &
  v_L \in [10^{-6},10^2] \, \text{GeV},\quad v_{\sigma} \in [10^3,1.2 \times 10^4] \, \text{GeV}.
\end{align}
This first general scan is useful to see the impact of the stellar
cooling constraint, Eq.~(\ref{eq:60}), on the parameter space. In
fact, to comply more easily with this constraint we have sampled
$v_\sigma$ values above 1~TeV, aware that lower values could be
possible. 
The resulting allowed region in the ($v_\sigma,v_L$) plane is shown in
Fig.~\ref{fig:stellar}. The yellow region is the set of values of
$(v_\sigma,v_L)$ that satisfy Eq.~(\ref{eq:60}). From here we
immediately see that the allowed range of $v_L$ is much smaller than
what we start with. This requires a stringent restriction on $v_L$,
i.e. $v_L < 0.5$ GeV. As mentioned above, this limit may be avoided in
the presence of explicit soft lepton number violation terms. We took
this in account to optimize the constrained scans below. 
Note that Eq.~(\ref{eq:60}) is a relation among three vevs, but there is another relation coming from the $W$ mass,
\begin{equation}
\label{eq:59a}
M_W=\frac{1}{2}\, g\, \sqrt{v_L^2+v_\phi^2},
\end{equation}
which explains the boundary in the plane ($v_\sigma,v_L$). In
Fig.~\ref{fig:stellar} we also plot the allowed points once all the
constraints on the model are implemented. The color code is as
follows: the points in dark green have $v_L> 0.1$ GeV, those in green
have $v_L\in[0.01,01]$GeV and those in yellow green have $v_L < 0.01$
GeV. We have imposed a lower bound on the value of $v_\sigma$ which is
visible in the figure.  

\subsection{Constrained Scan}

\begin{figure}[htb]
  \centering
  \includegraphics[width=0.45\textwidth]{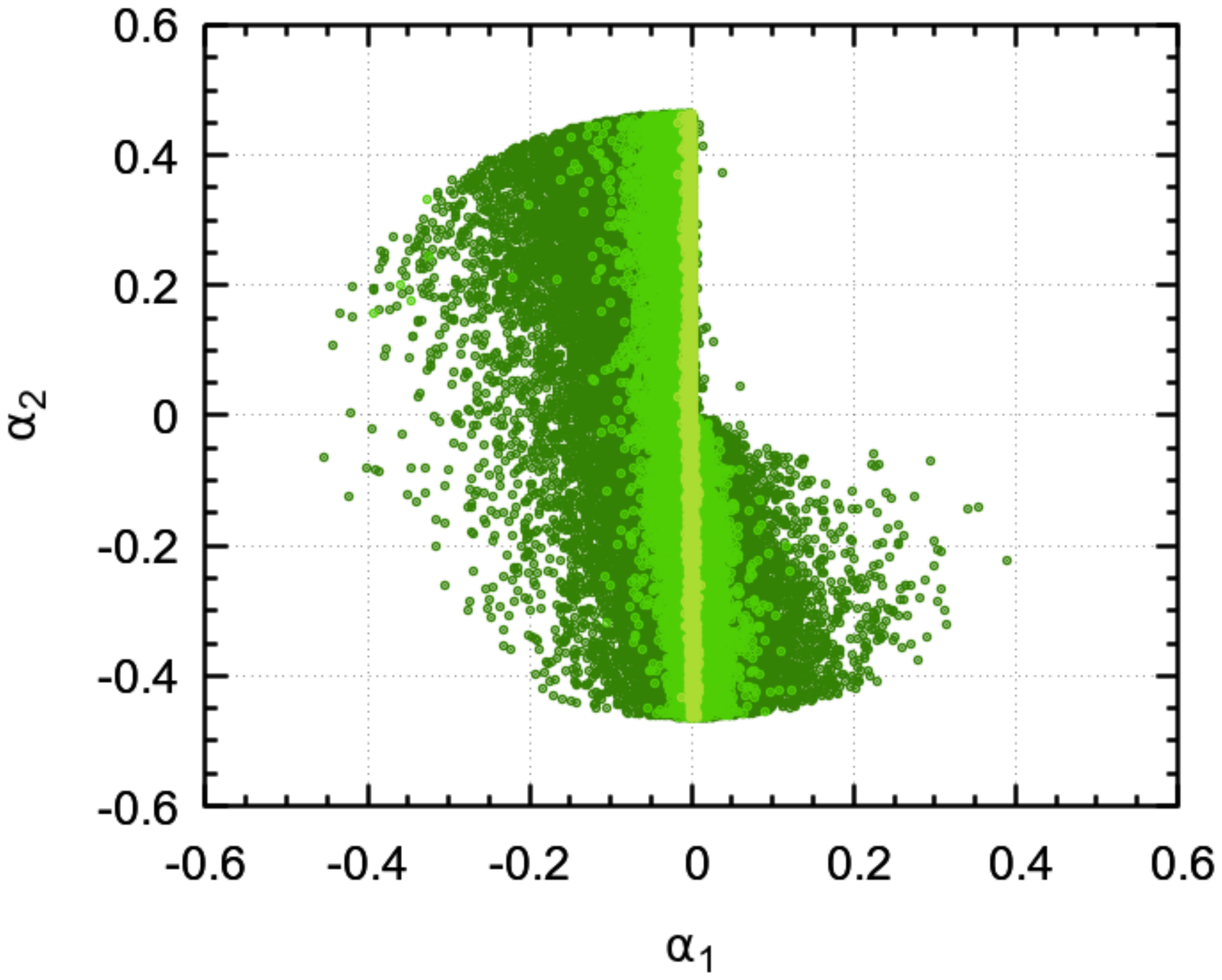}
  \includegraphics[width=0.45\textwidth]{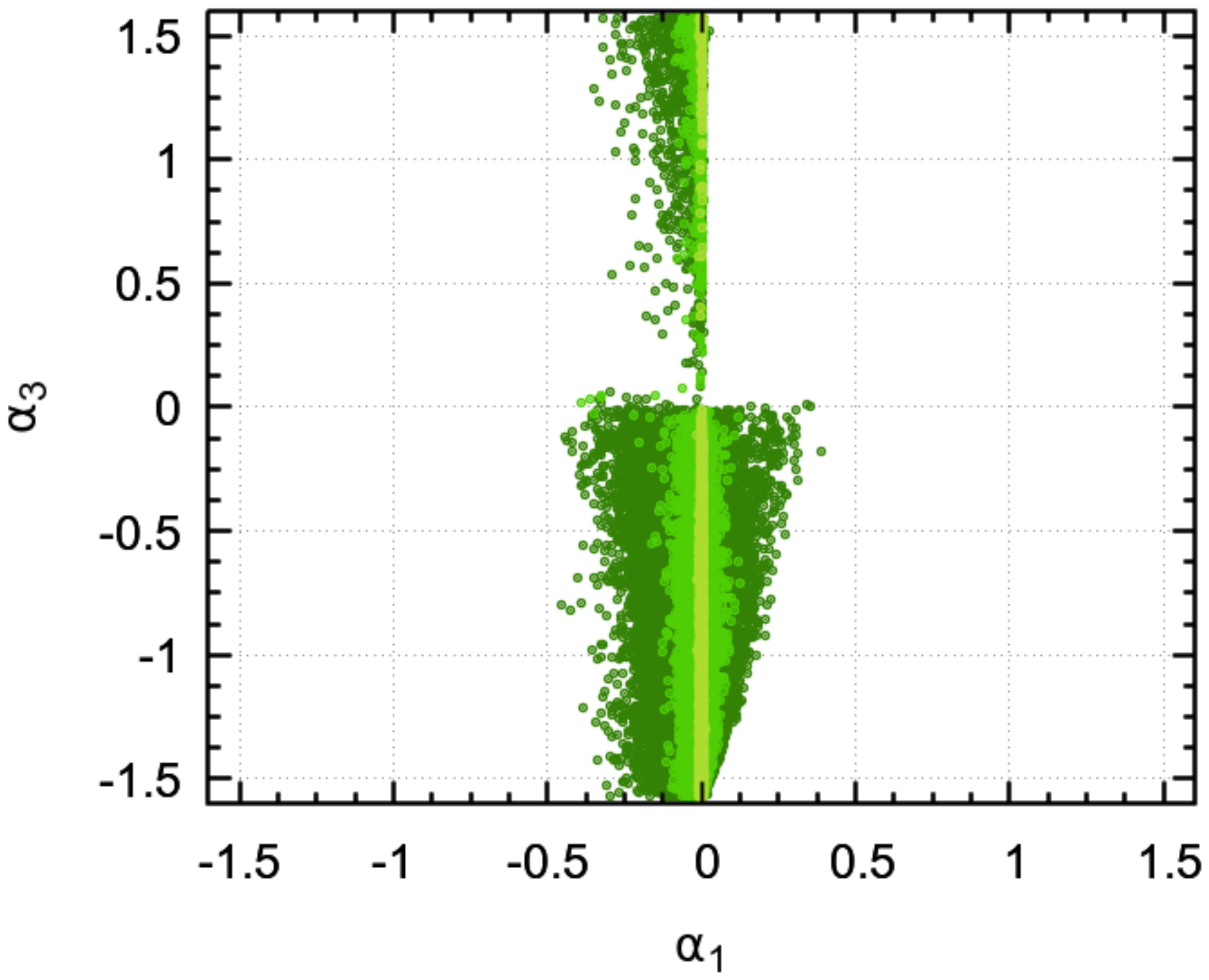}
  \caption{$\alpha_2$ and $\alpha_3$ versus $\alpha_1$. The color code is as in Fig.~\ref{fig:stellar}. See text for the explanation.}
  \label{fig:alfas}
\end{figure}

\begin{figure}[htb]
  \centering
   \includegraphics[width=0.45\textwidth]{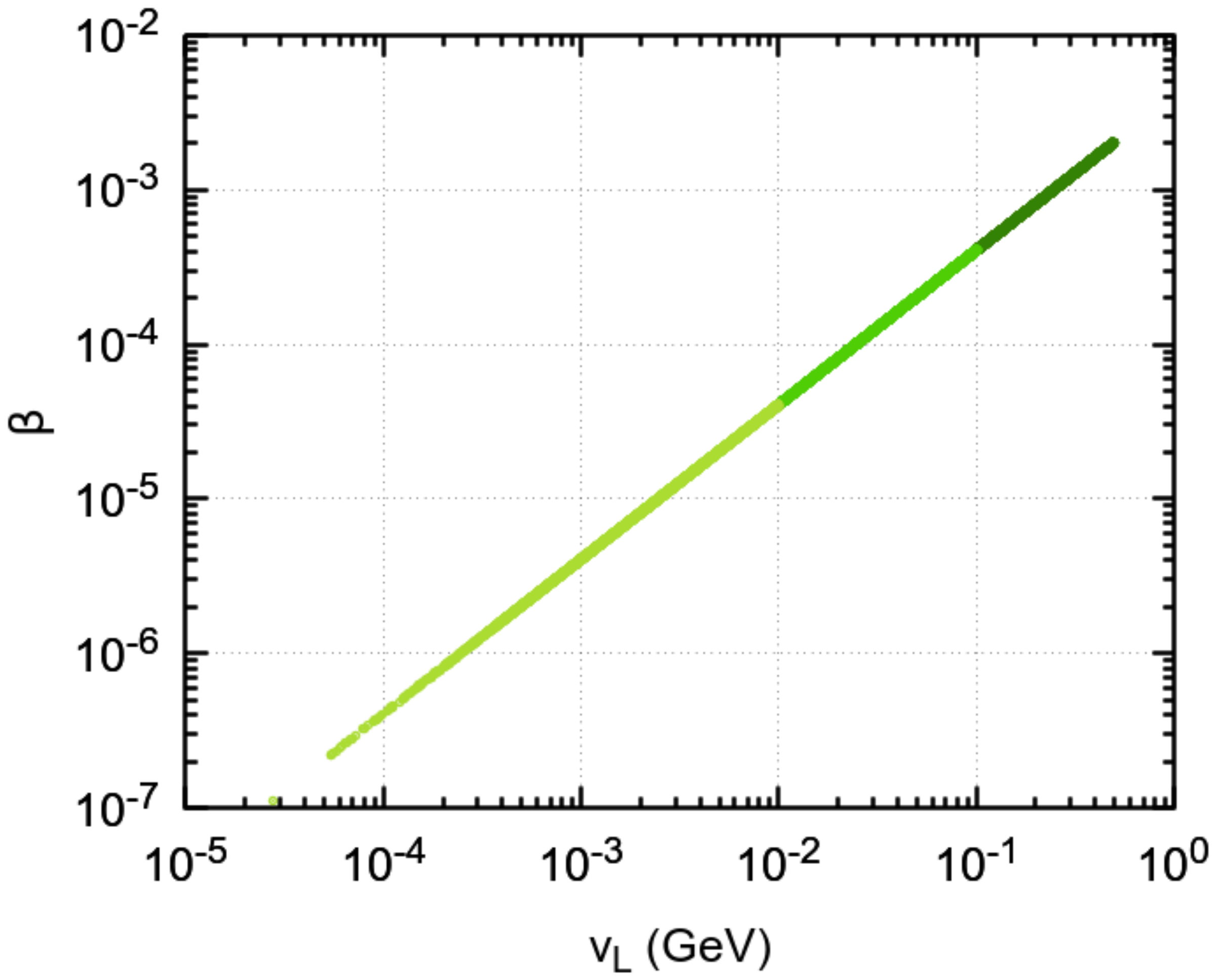}
  \includegraphics[width=0.45\textwidth]{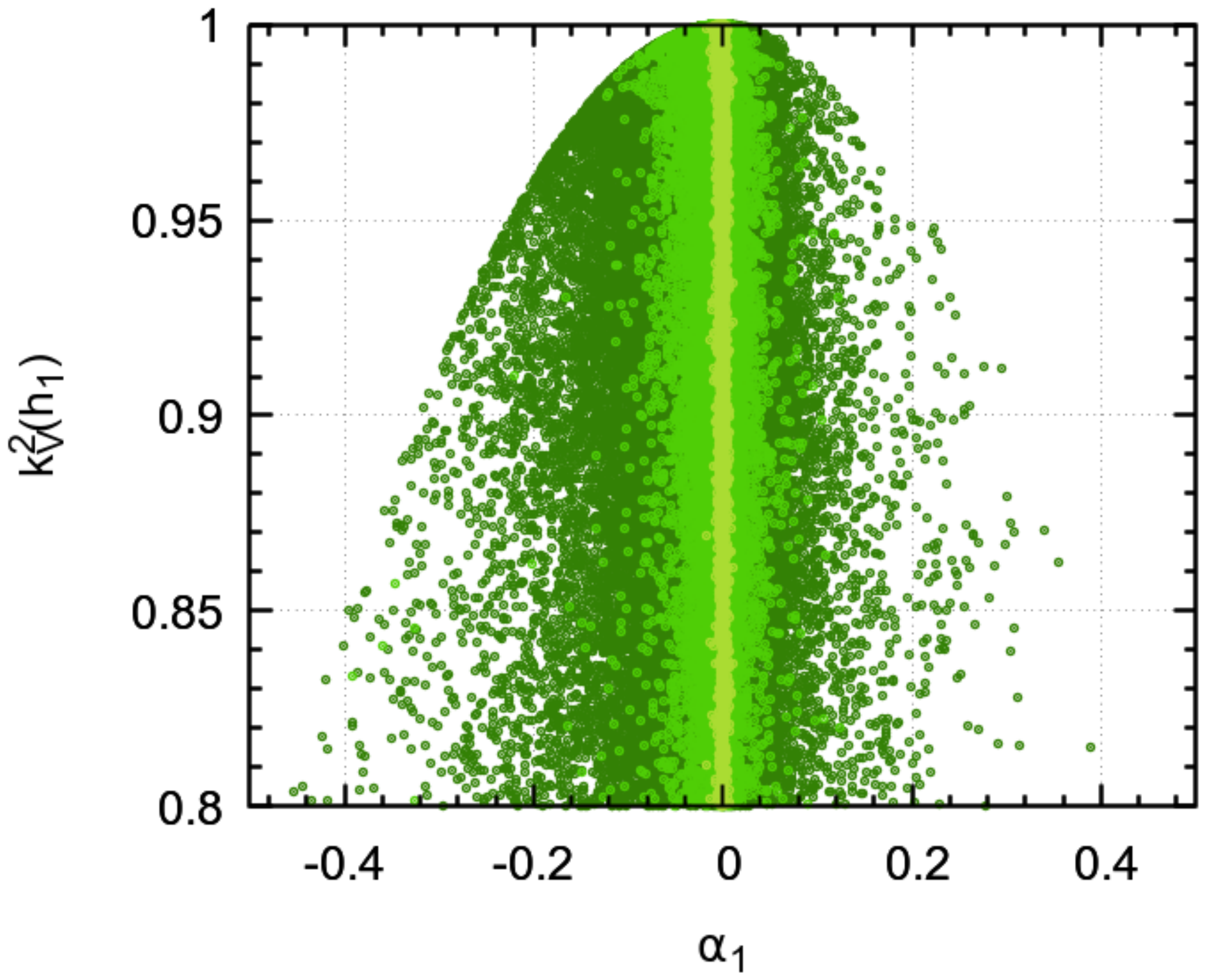} 
  \caption{Left panel: Correlation between $\beta$ and $v_L$. Right
    panel: Correlation between $k_V^2(h_1)$ and $\alpha_1$. The color
    code is as in Fig..~\ref{fig:stellar}. See text for the
    explanation.} 
  \label{fig:betaalpha}
\end{figure}

Using the above result, we have performed a constrained scan where all points satisfy the astrophysical constraint, Eq.~(\ref{eq:60}). 
We have also required that the points satisfy the LHC constraints on the signal strengths given in Table~\ref{tab:1} and Table~\ref{tab:2}, at the 3$\sigma$ level. 
As explained above, we have separated all the points in three bins, according to the value of the vev $v_L$ as follows:
\begin{align}
  \label{eq:63}
  &v_L > 0.1\text{GeV} && \text{dark green points},&&\nonumber\\
  &v_L \in [0.01,0.1] \text{GeV} && \text{green points},&&\\
  &v_L < 0.01\text{GeV} && \text{yellow green points}. &&\nonumber
\end{align}
We have further restricted the scan to scalar masses values below 600 GeV, which might be explored in the next generation of collider experiments.
How the allowed model parameters are constrained is shown in Fig.~\ref{fig:alfas} and Fig.~\ref{fig:betaalpha}. 
The mixing angles in the neutral scalar rotation matrix are shown in Fig.~\ref{fig:alfas}.
Their values are restricted by the LHC constraint on the signal strengths, which we imposed at the 3$\sigma$ level. The color code is defined in
Eq.~(\ref{eq:63}). We see from Eq.~(\ref{eq:23b}) that smaller values of $v_L$ require the the numerator of that equation to be small, otherwise perturbative unitarity on $\lambda_L$ would be violated. One can verify that the smallness of the numerator occurs for 
 $\alpha_1$  close to zero and for a compressed spectrum, as we will discuss below.
We should also note that, as the allowed values of $v_L$ are quite small, $\beta$ in Eq.~\ref{eq:17}, is a very small angle, as shown in Fig.~\ref{fig:betaalpha}; here,
we show in the left panel the relation of $\beta$ with $v_L$, and in
the right panel the correlation between $k_V^2(h_1)$ and
$\alpha_1$. The color code is the same. We see that the lower limit on
$k_V^2(h_1)$, set in Eq.~(\ref{eq:57}), corresponds indeed to the signal
strengths at 3$\sigma$, as we have points for all values of
$k_V^2(h_1)$ in that range. If we enforced the signal strength
constraints at 2$\sigma$, the range would be reduced as we will show
below. 

\subsection{Compressed Higgs Boson Spectra}
  
We have found that, in order to ensure very small values of the lepton number breaking vev $v_L$, the spectrum of Higgs bosons tends to be compressed. 
The reason can be traced to the perturbative unitarity constraints on the quartic coulings, in particular those on $\lambda_L$. This can be easily understood by considering Eq.~(\ref{eq:23b}). As $\lambda_L$ is inversely proportional to the third power of $v_L$, for small values of $v_L$ the numerator must be very small, to prevent $\lambda_L$ violating the perturbative unitarity constraints. 
It turns out that this  is achieved by a compressed spectrum and a value of the mixing angle $\alpha_1$ close to zero. In fact, one can show that, for $v_L\ll v_\phi$, the numerator of Eq.~(\ref{eq:23b}) vanishes for $\alpha_1=\alpha_3=0$ and for equal masses. We did not impose these last conditions on our scan, it just turned 
out the that the good points have this profile.   
\begin{figure}[htb]
  \centering
   \includegraphics[width=0.31\textwidth]{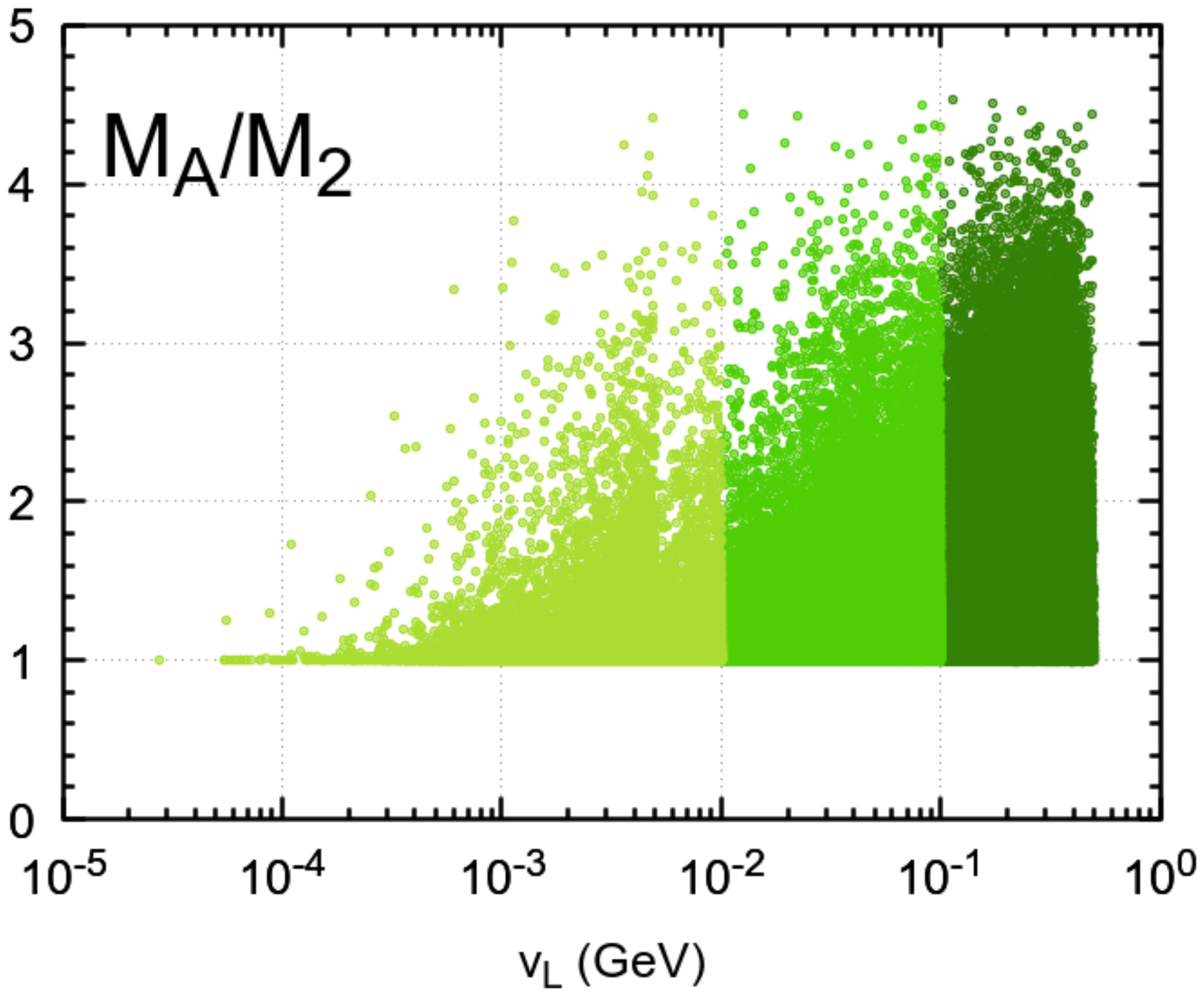}
  \includegraphics[width=0.31\textwidth]{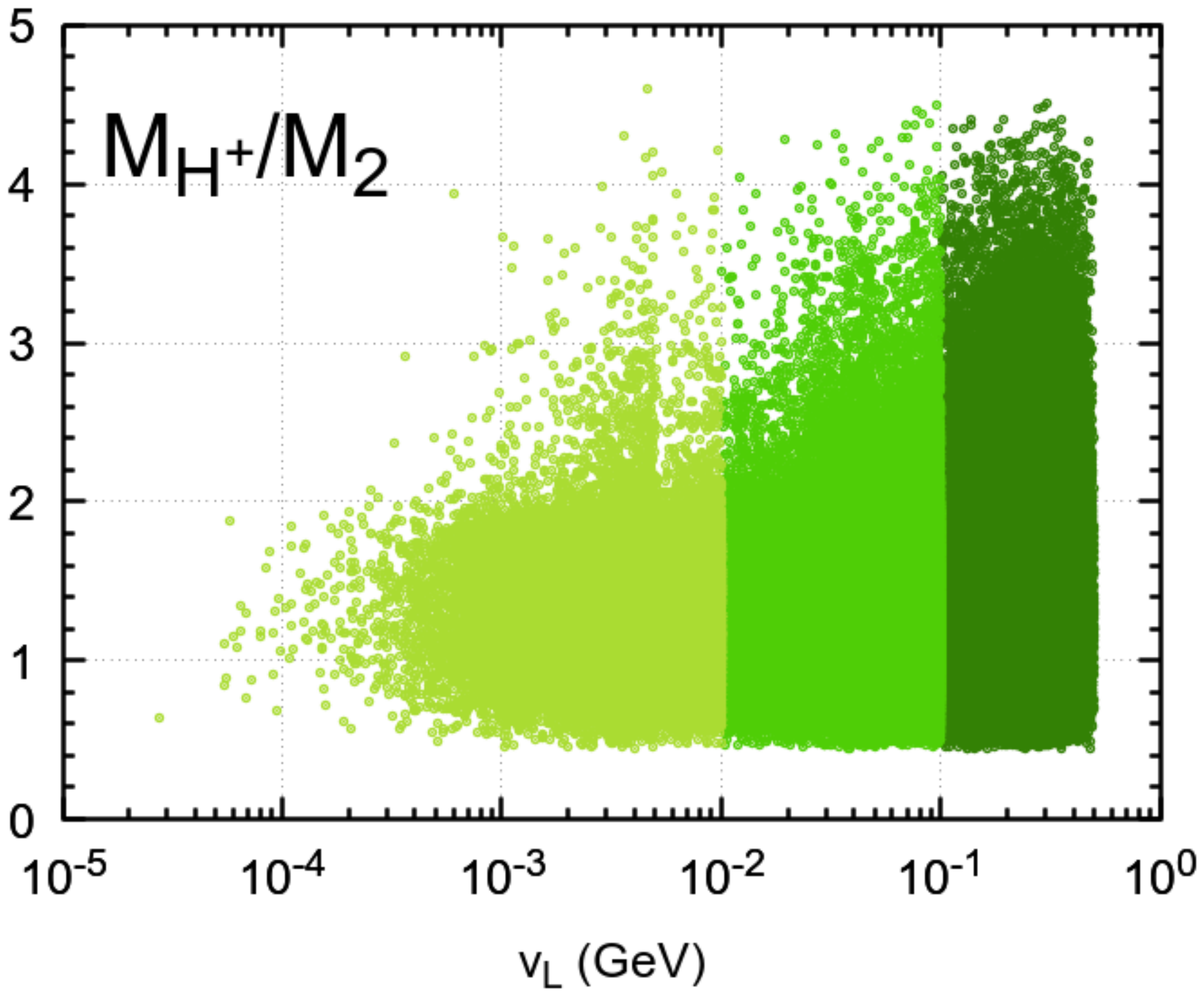} 
 \includegraphics[width=0.31\textwidth]{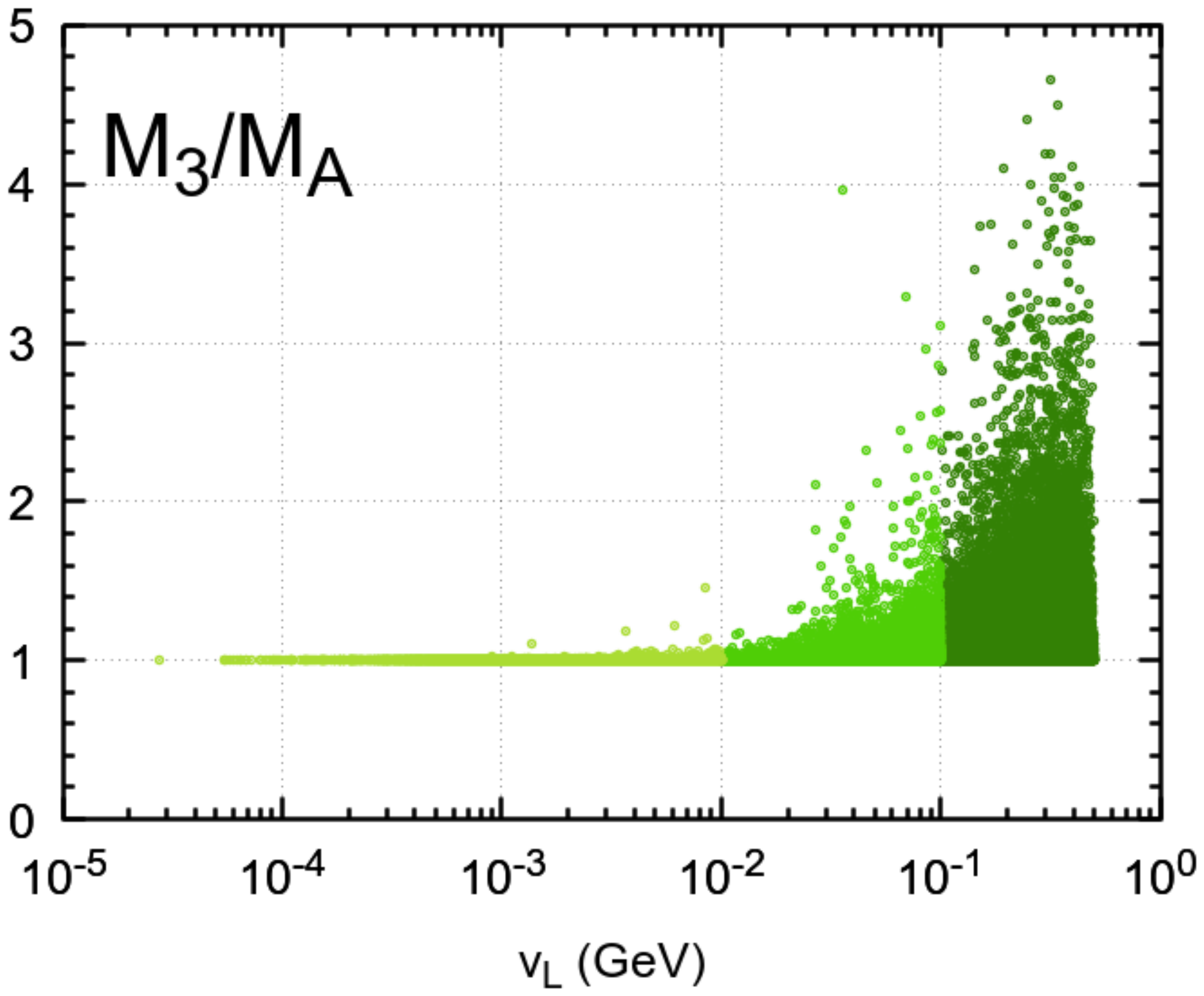}
  \caption{Correlation between $M_{A}/M_{2}$ (left), $M_{H^+}/M_{2}$ (middle), $M_{3}/M_{A}$ (right) and $v_L$. The color code as in Fig.~\ref{fig:stellar}. See text for
    explanation.}
  \label{fig:Ratios1}
\end{figure}
This is shown in the plots in Fig.~\ref{fig:Ratios1}. Notice that this compression is much stronger than that usually imposed by the oblique parameters. One can see that, for larger values of $v_L$, the points satisfying all the constraints, including those from the oblique parameters, can have a sizeable splitting.

\section{Invisible Higgs decays at the LHC}
\label{sec:invis-higgs-decays}

It has long been noticed that theories of neutrino mass where spontaneous violation of ungauged lepton number symmetry takes place at collider-accessible scales all lead 
to the phenomenon of invisibling Higgs decay bosons~\cite{Joshipura:1992hp}.
Collider implications have been widely discussed in the literature in various theory contexts and collider setups~\cite{Romao:1992zx,Eboli:1994bm,DeCampos:1994fi,Romao:1992dc,deCampos:1995ten,deCampos:1996bg,Diaz:1998zg,Hirsch:2004rw,Hirsch:2005wd,Bonilla:2015uwa,Bonilla:2015jdf}.

In our model, the new decay channels of the CP-even scalars that contribute to the invisible decays are just $h_{i}\to JJ$ and
$h_{i}\to 2h_{j}$ (when $M_{i}> 2 M_{j}$). The latter also contributes as $h_{i}\to 2h_{j}\to 4J$.
The Higgs-majoron coupling is given by (no summation on $a$),
\begin{equation}
\label{ghJJ}
g_{h_{a}JJ}=-
\left(\frac{(\mathcal{O}^{I}_{21})^2}{v_{\phi}}\mathcal{O}^{R}_{a1}+
            \frac{(\mathcal{O}^{I}_{22})^2}{v_{L}}\mathcal{O}^{R}_{a2}+
            \frac{(\mathcal{O}^{I}_{23})^2}{v_{\sigma}}\mathcal{O}^{R}_{a2}
            \right) M_{a}^2,
\end{equation}
where $\mathcal{O}^{I}_{ij}$ are the elements of the rotation matrix in Eq.~(\ref{eq:26}), and the decay width is given by
\begin{equation}
  \label{eq:13}
\Gamma(h_a \to JJ) = \frac{1}{32\pi} \frac{g^2_{h_{a} JJ}}{M_a}
\, .
\end{equation}
Following our conventions, the trilinear coupling $h_{2}h_{1}h_{1}$ are giving by:
\begin{small}
\begin{equation}
\begin{split}
g_{h_{2}h_{1}h_{1}}= - & \bigg( 6 \, {\lambda} \,
({\mathcal{O}^R_{11}})^2 \, {\mathcal{O}^R_{21}} \, {v_{\phi }} +
{\beta_5} \, ({\mathcal{O}^R_{12}})^2 \, {\mathcal{O}^R_{21}} \,
{v_{\phi }} + {\beta_2} \, ({\mathcal{O}^R_{13}})^2 \,
{\mathcal{O}^R_{21}} \, {v_{\phi }} + 2 \, {\beta_5} \,
{\mathcal{O}^R_{11}} \, {\mathcal{O}^R_{12}} \, {\mathcal{O}^R_{22}}
\, {v_{\phi }}\\ & - {\beta_4} \, ({\mathcal{O}^R_{13}})^2 \,
{\mathcal{O}^R_{22}} \, {v_{\phi }}  
+ 2 \, {\beta_2} \, {\mathcal{O}^R_{11}} \, {\mathcal{O}^R_{13}} \,
{\mathcal{O}^R_{23}} \, {v_{\phi }} - 2 \, {\beta_4} \,
{\mathcal{O}^R_{12}} \, {\mathcal{O}^R_{13}} \, {\mathcal{O}^R_{23}}
\, {v_{\phi }} + 2 \, {\beta_5} \, {\mathcal{O}^R_{11}} \,
{\mathcal{O}^R_{12}} \, {\mathcal{O}^R_{21}} \, {v_L}\\
& - {\beta_4} \,({\mathcal{O}^R_{13}})^2 \, {\mathcal{O}^R_{21}} \,
{v_L} + {\beta_5} 
\, ({\mathcal{O}^R_{11}})^2 \, {\mathcal{O}^R_{22}} \, {v_L}
+ 6 \, {\lambda_L} \, ({\mathcal{O}^R_{12}})^2 \,
{\mathcal{O}^R_{22}} \, {v_L} + {\beta_3} \, ({\mathcal{O}^R_{13}})^2
\, {\mathcal{O}^R_{22}} \, {v_L}\\
&- 2 \, {\beta_4} \,
{\mathcal{O}^R_{11}} \, {\mathcal{O}^R_{13}} \, {\mathcal{O}^R_{23}}
\, {v_L} + 2 \, {\beta_3} \, {\mathcal{O}^R_{12}} \,
{\mathcal{O}^R_{13}} \, {\mathcal{O}^R_{23}} \, {v_L} 
 + {\beta_1} \,  \left( ({\mathcal{O}^R_{12}})^2 \,
  {\mathcal{O}^R_{21}} \, {v_{\phi }} + ({\mathcal{O}^R_{11}})^2 \,
  {\mathcal{O}^R_{22}} \, {v_L} \right.\\
&\left.
+ 2 \, {\mathcal{O}^R_{11}} \,
  {\mathcal{O}^R_{12}} \,  \left( {\mathcal{O}^R_{22}} \, {v_{\phi }}
    + {\mathcal{O}^R_{21}} \, {v_L} \right)  \right) 
+ 2 \, {\beta_2} \, {\mathcal{O}^R_{11}} \, {\mathcal{O}^R_{13}} \,
{\mathcal{O}^R_{21}} \, {w} - 2 \, {\beta_4} \, {\mathcal{O}^R_{12}}
\, {\mathcal{O}^R_{13}} \, {\mathcal{O}^R_{21}} \, {w}\\
&- 2 \, {\beta_4} \, {\mathcal{O}^R_{11}} \, {\mathcal{O}^R_{13}} \,
{\mathcal{O}^R_{22}} \, {w} + 2 \, {\beta_3} \, {\mathcal{O}^R_{12}}
\, {\mathcal{O}^R_{13}} \, {\mathcal{O}^R_{22}} \, {w} 
+ {\beta_2} \, ({\mathcal{O}^R_{11}})^2 \, {\mathcal{O}^R_{23}} \,
{w} - 2 \, {\beta_4} \, {\mathcal{O}^R_{11}} \, {\mathcal{O}^R_{12}}
\, {\mathcal{O}^R_{23}} \, {w}\\
&+ {\beta_3} \, ({\mathcal{O}^R_{12}})^2
\, {\mathcal{O}^R_{23}} \, {w} + 6 \, {\lambda_{\sigma }} \,
({\mathcal{O}^R_{13}})^2 \, {\mathcal{O}^R_{23}} \, {w} \bigg), 
\end{split}
\end{equation}
\end{small}
and hence, for example when $2 M_1 < M_2$, we have the decay width $h_{2}\to h_{1}h_{1}$ given by
\begin{equation}
\Gamma\left(h_{2}\to h_{1}h_{1}\right)=\frac{g_{h_{2}h_{1}h_{1}}^{2}}
{32 \pi M_2}\left(1-\frac{4 M_1^{2}}{M_2^{2}}\right)^{1/2}.
\end{equation}
We have computed all the decay channels of the neutral and charged scalars in order to obtain their branching fractions. For finding the new Feynman rules and computing the amplitudes and decay rates we used the new software \texttt{FeynMaster}~\cite{Fontes:2019wqh}, that makes use of \texttt{FeynRules}~\cite{Christensen:2008py}
\texttt{QGRAF}~\cite{Nogueira:1991ex},
and \texttt{FeynCalc}~\cite{Mertig:1990an, Shtabovenko:2016sxi}.
For the computation of the decay widths $h\to \gamma\gamma$ and $h\to Z\gamma$, we used the expressions and conventions given in Ref.~\cite{Fontes:2014xva}.

\begin{figure}[htb]
  \centering
  \includegraphics[width=0.45\textwidth]{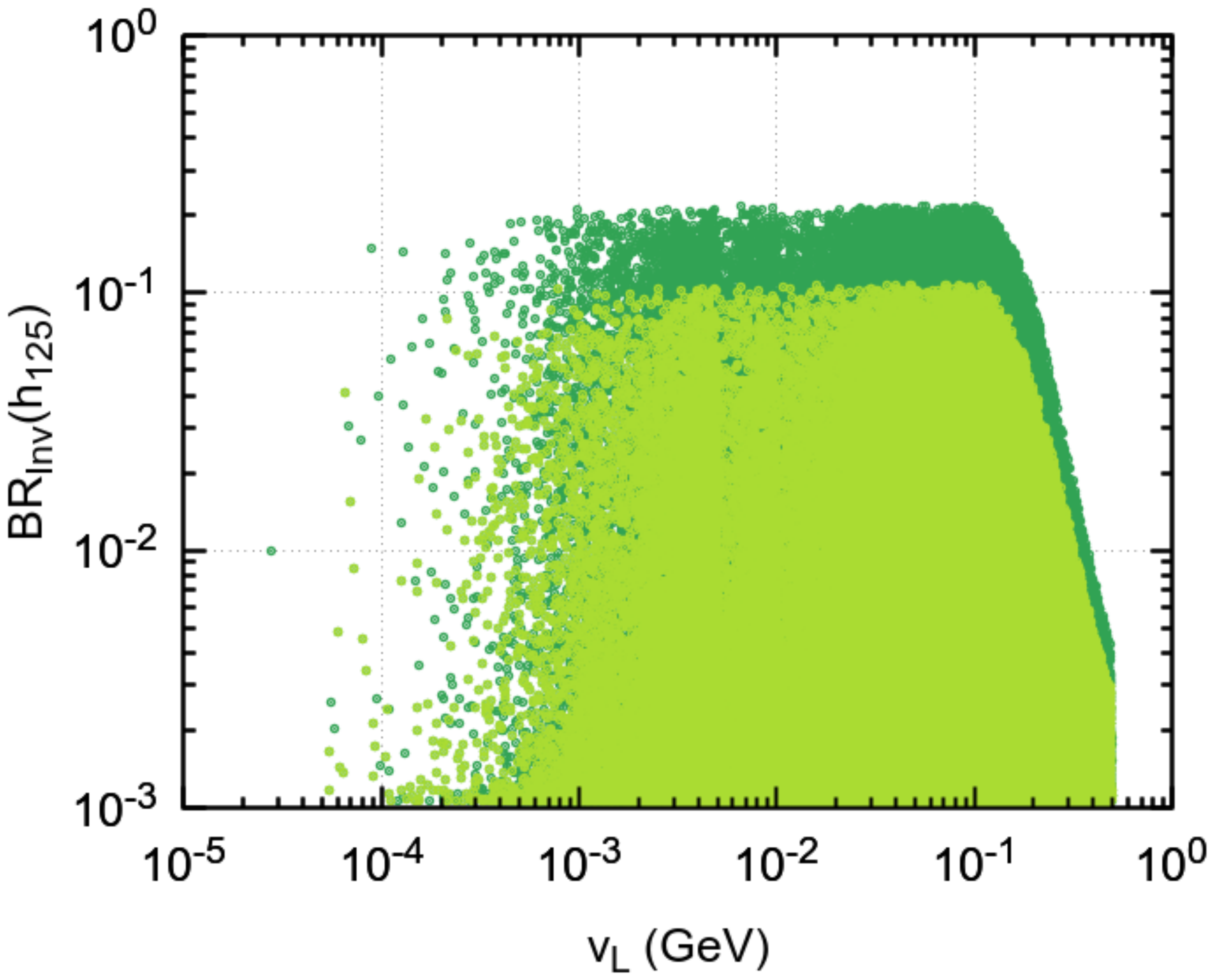}
    \includegraphics[width=0.45\textwidth]{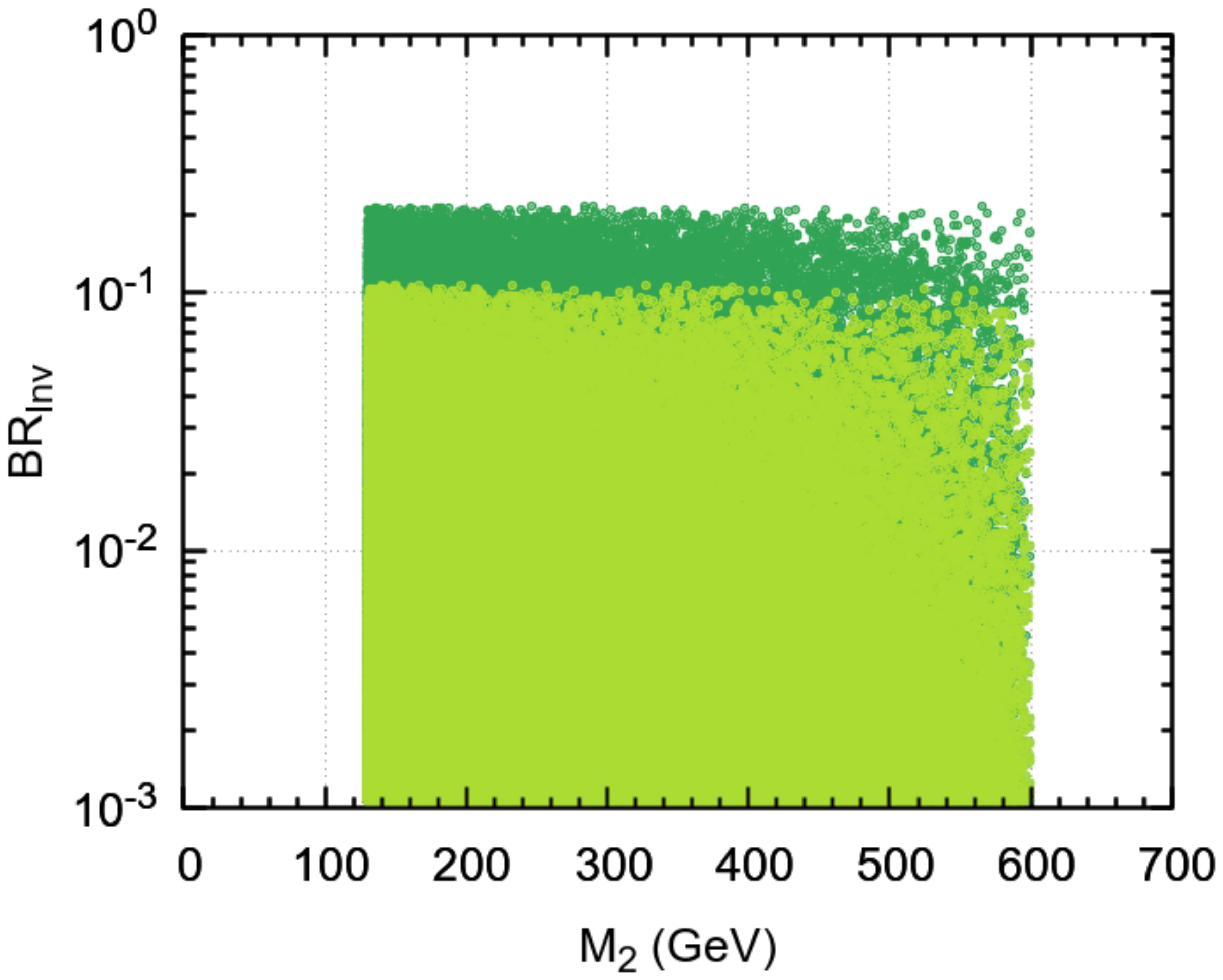}
    \caption{Left panel: BR${}_{\rm inv}$ versus $v_L$. Right panel:
      BR${}_{\rm inv}$ versus  $M_2$. In dark green the LHC constraints are imposed at 3$\sigma$, while in yellow green is shown the result of requiring 
     2$\sigma$ LHC constraints.}
    \label{fig:BRinv}
\end{figure}

The results obtained for the invisible branching ratios of the 125 GeV Higgs boson are shown in Fig.~\ref{fig:BRinv}.  In the left panel we show the BR($h_{125}\to$ 
invisible) as a function of $v_L$ and in the right panel as a function of the mass of the second scalar Higgs boson.  
The points in dark green satisfy the LHC constraints on the signals strengths at 3$\sigma$ level. The points in yellow green satisfy those constraints at 2$\sigma$. 

\begin{figure}[htb]
  \centering
  \includegraphics[width=0.45\textwidth]{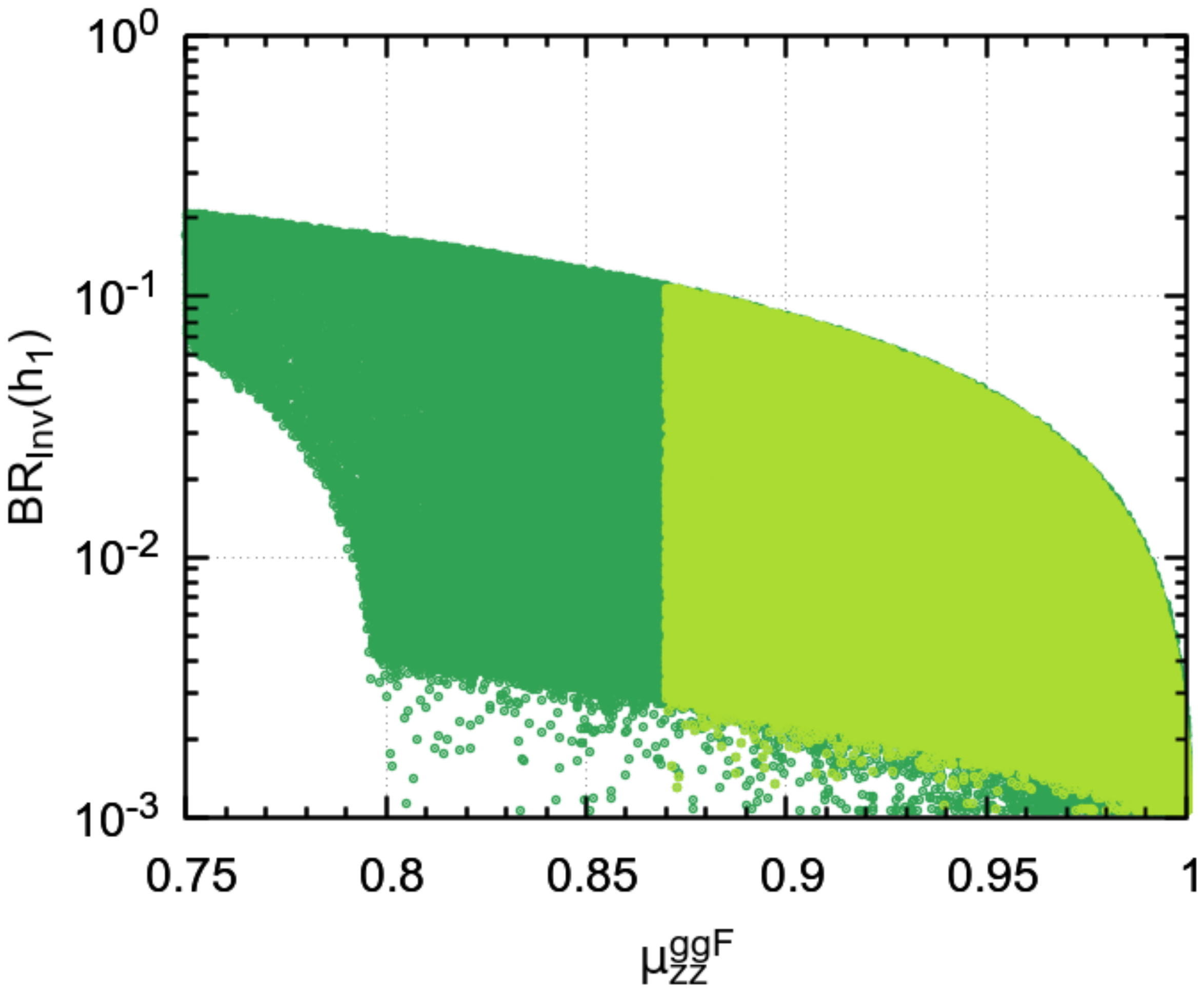}
  \includegraphics[width=0.45\textwidth]{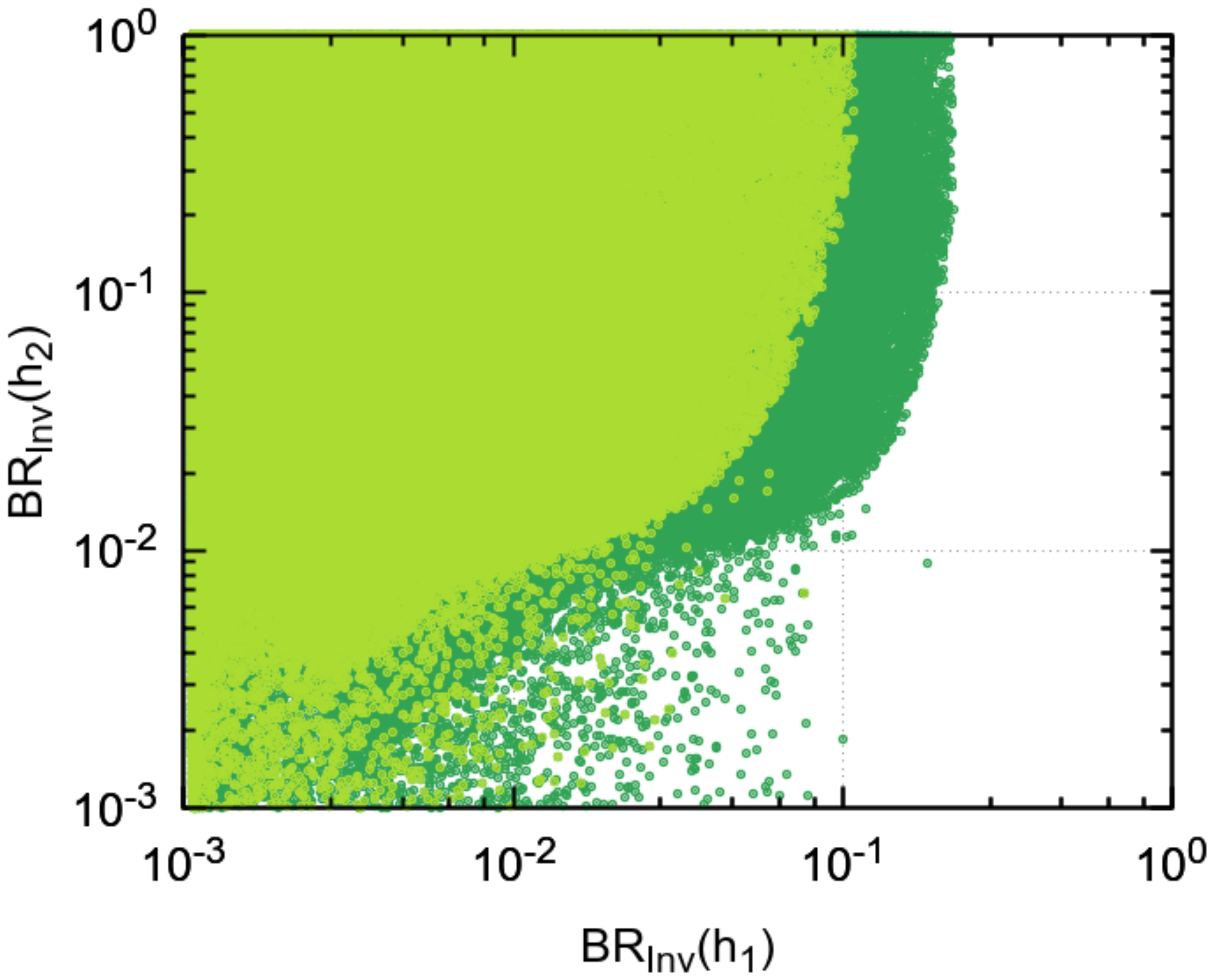}
    \caption{Left panel: BR${}_{\rm inv}$ versus $\mu_{zz}^{ggF}$. Right panel: BR${}_{\rm inv}(h_2)$ versus BR${}_{\rm inv}(h_1)$. In dark green the LHC constraints are imposed at 3$\sigma$, while in yellow green is shown the result of 
     requiring 2$\sigma$ LHC constraints.}
    \label{fig:BRinvb}
\end{figure}

We see that an invisible branching ratio around 20\%, close to the present upper bound \cite{Sirunyan:2018owy,Aaboud:2019rtt}, is possible within this model. This is consistent with all the LHC constraints including those on the signal strengths at the 3$\sigma$ level. However, if 2$\sigma$ limits are applied, the ratio reduces to a maximum of 10\%. This is better illustrated on the left panel of Fig.~\ref{fig:BRinvb}, where we plot the invisible branching ratio of the 125 GeV Higgs boson, versus the signal strength of the Higgs boson produced via gluon fusion and decaying in the $ZZ$ final state. 

\begin{figure}[htb]
	\centering
	\includegraphics[width=0.45\textwidth]{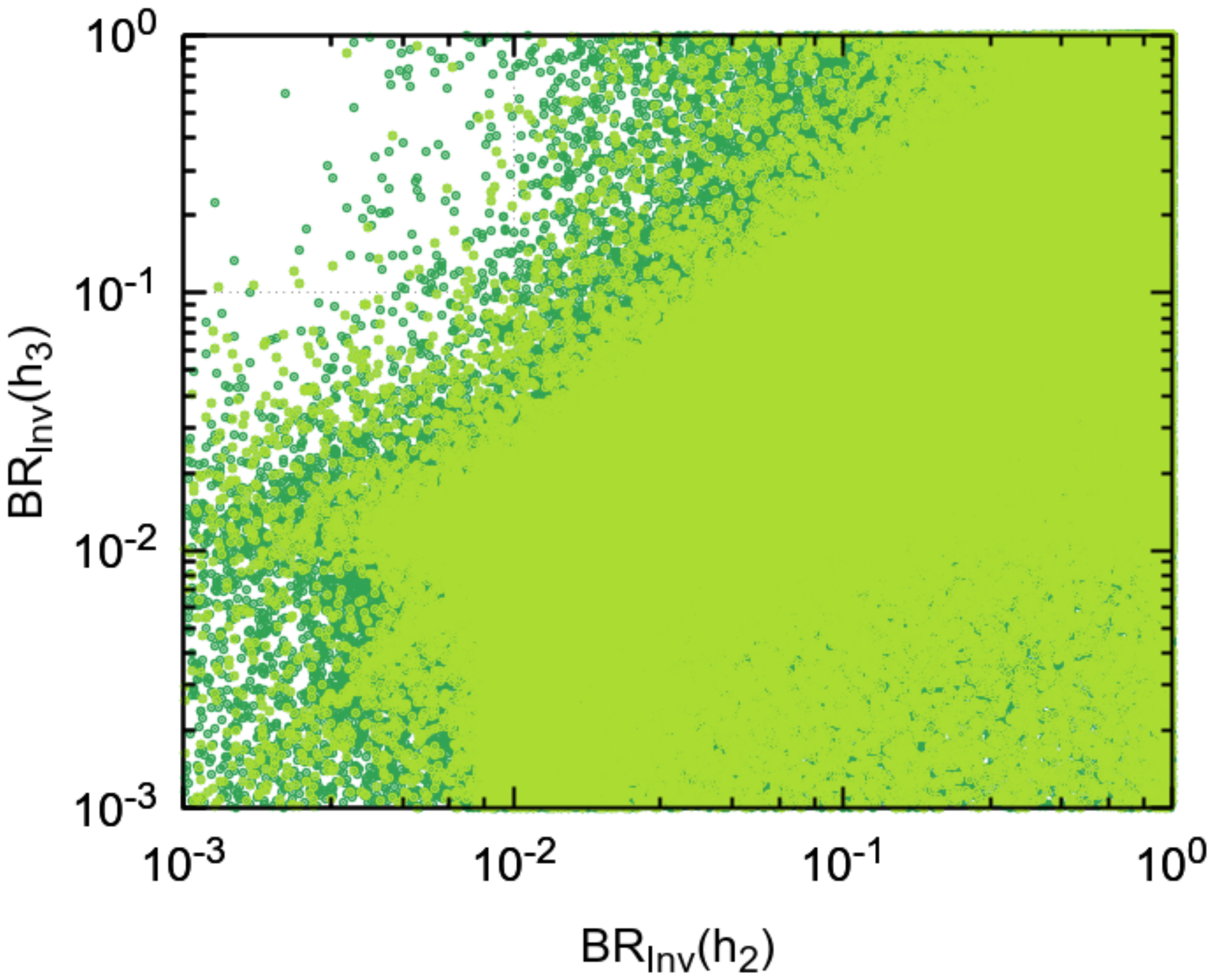}
	\includegraphics[width=0.45\textwidth]{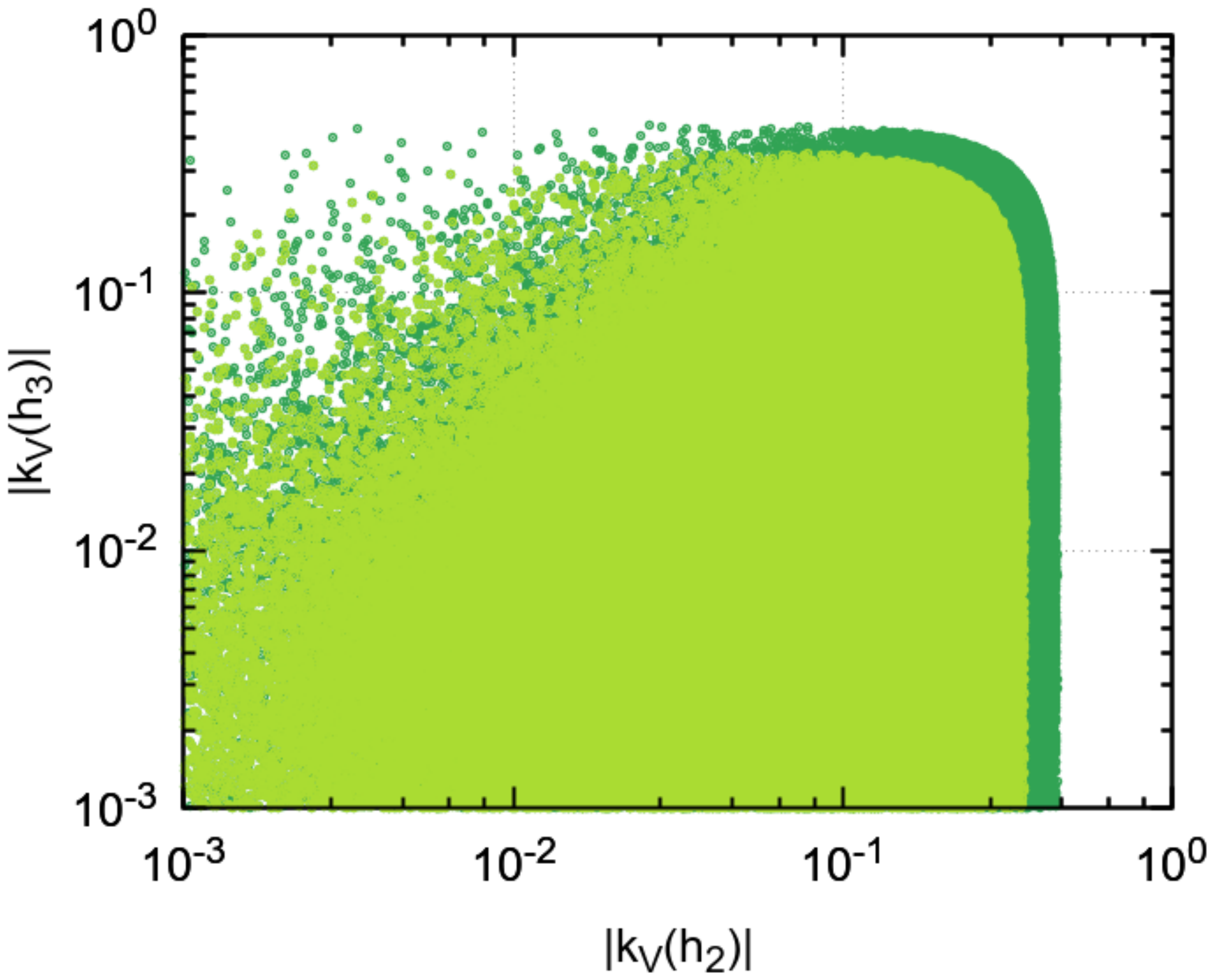}
	\caption{Left panel: BR${}_{\rm inv}(h_3)$ versus BR${}_{\rm
            inv}(h_2)$. Right panel: The effective couplings to vector
          bosons with respect to that of the SM, $k_V(h_i)$ of the second
and third CP even Higgs boson. Notice that these couplings are bounded
due to the sum rule of Eq.~(\ref{eq:3a}). In dark green the LHC constraints are
imposed at 3$\sigma$, while in yellow green (on top) is shown the
result of requiring 2$\sigma$ LHC constraints.} 
	\label{fig:BRinvb2}
\end{figure}

If it is found at the LHC that this signal strength becomes closer to one, then the invisible branching ratio in this model will be smaller, which can be important for LHC searches. On the right panel of Fig.(\ref{fig:BRinvb}) we show the correlation between the invisible branching ratios of the two lightest CP even Higgs bosons. 
We see that the invisible branching ratio of the Higgs boson lying above the one with $m_H=125$GeV can have a wide range of values, from very small (hence visible) to close to an 100\% invisible branching ratio. To better understand the different possibilities, we have also plotted in the left panel of Fig.~\ref{fig:BRinvb2} the correlation between the invisible branching ratios of the second and third CP even Higgs bosons. 
On the right panel of Fig.~\ref{fig:BRinvb2} we show the absolute
value of their couplings to vector bosons, which gives a measure of the
probability that they are produced and observed.
We notice that the effective couplings of the second and third CP
  even Higgs boson to vector bosons are limited to a maximum value
  below 1, due to the sum rule in Eq.~(\ref{eq:3a}) and to the fact that
  the coupling of the lightest Higgs boson is
  bounded from below to be in agreement with LHC data, as shown in
  Eq.~(\ref{eq:57}).
From these figures we
see that the model has a very rich structure with many
possibilities. In order to better illustrate this we selected three
benchmark points with quite different characteristics.

\subsection{\textbf{P1}: BR${}_{\rm inv}(h_1)>0.1$, BR${}_{\rm inv}(h_2)<0.01$, BR${}_{\rm inv}(h_3)<0.01$}

These points typically have $2 M_{H^+} < M_{h_2}, M_{h_3}$, thus allowing the
second and third CP even Higgs bosons to decay visibly into the charged
one. One such example is given in Table~\ref{tab:P1}.  
\begin{table}[htb]\small
\begin{tabular}{|c|c|c|c|c|c|c|c|c|c|c|c|}\hline
  $M_{h_{1}}$&$M_{h_{2}}$&$M_{h_{3}}$&$M_{H^{+}}$&$M_{A}$&$v_L$
  &BR${}_{\rm inv}$
  &BR${}_{\rm inv}$&BR${}_{\rm  inv}$&$|k_V|$&$|k_V|$&$|k_V|$
  \\
  \!(GeV)\!&\!(GeV)\!&\!(GeV)\!&\!(GeV)\!&\!(GeV)\!&\!(GeV)\!
  &$h_1$&$h_2$&$h_3$&$h_1$&$h_2$&$h_3$\\\hline\hline
  125.0&298.0&301.4&133.2&299.0&0.13&$0.11$&$\!\!5\!\times\!\! 10^{-5}\!\!$
  &$\!\!5\!\times\!\! 10^{-5}\!\!$&$0.95$&$0.16$&$0.26$\\\hline
\end{tabular}
\caption{Parameters for point P1.}
\label{tab:P1}
\end{table}

We see that all the CP even Higgs bosons have non-negligible couplings to
vector bosons, although these are bounded due to the sum rule of
Eq.~(\ref{eq:3a}). In contrast, only the lightest one (the 125 GeV SM Higgs) has an observable invisible branching ratio.

\subsection{\textbf{P2}: BR${}_{\rm inv}(h_1)>0.1$, BR${}_{\rm inv}(h_2)>0.1$, BR${}_{\rm inv}(h_3)<0.01$}

These points typically have $2 M_{H^+} > M_{h_2}, M_{h_3}$, so that
the second and third CP even Higgs bosons do not decay visibly into the
charged one. One such example is given in Table~\ref{tab:P2}.
\begin{table}[htb]\small
\begin{tabular}{|c|c|c|c|c|c|c|c|c|c|c|c|}\hline
  $M_{h_{1}}$&$M_{h_{2}}$&$M_{h_{3}}$&$M_{H^{+}}$&$M_{A}$&$v_L$
  &BR${}_{\rm inv}$
  &BR${}_{\rm inv}$&BR${}_{\rm  inv}$&$|k_V|$&$|k_V|$&$|k_V|$
  \\
  \!(GeV)\!&\!(GeV)\!&\!(GeV)\!&\!(GeV)\!&\!(GeV)\!&\!(GeV)\!&$h_1$&$h_2$&$h_3$
  &$h_1$&$h_2$&$h_3$\\\hline\hline 
  125.0&251.9&277.7&167.8&276.7&$0.13$&$0.18$&$0.13$
  &$\!2\!\times\! 10^{-3}\!$&$0.95$&$0.31$&$0.05$\\\hline
\end{tabular}
\caption{Parameters for point P2.}
\label{tab:P2}
\end{table}

We see that the second CP even Higgs boson can have a sizable
invisible branching ratio, while its coupling to the vector bosons is
still large, therefore allowing, in principle, to be searched at the
LHC. In this case, due to Eq.~(\ref{eq:3a}), the coupling to vector
bosons of the third CP even Higgs boson is very small, making it very
difficult to produce.

\subsection{\textbf{P3}: BR${}_{\rm inv}(h_1)<0.01$, BR${}_{\rm inv}(h_2)<0.01$, BR${}_{\rm inv}(h_3)<0.01$}

These points have very small invisible branching ratios and non-negligible
couplings to the vector bosons, thus enhancing their visibility at the LHC.
Two such examples are given in Table~\ref{tab:P3}.  
\begin{table}[htb]\small
\begin{tabular}{|c|c|c|c|c|c|c|c|c|c|c|c|c|}\hline
 \!\!Point\!\!& $M_{h_{1}}$&$M_{h_{2}}$&$M_{h_{3}}$&$M_{H^{+}}$&$M_{A}$&$v_L$
  &BR${}_{\rm inv}$
  &BR${}_{\rm inv}$&BR${}_{\rm  inv}$&$\!|k_V|\!$&$\!|k_V|\!$&$\!|k_V|\!$
  \\
      &\!\!(GeV)\!\!&\!\!(GeV)\!\!&\!\!(GeV)\!\!&\!\!(GeV)\!\!
     &\!\!(GeV)\!\!&\!\!(GeV)\! 
    \!&$h_1$&$h_2$&$h_3$&$h_1$&$h_2$&$h_3$\\\hline\hline
  P3a&\!125.0\!&391.4&391.5&493.4&391.5&$\!\!6\!\!\times\! 10^{-2}\!\!\!$
  &$\!\!\!5\!\times\! 10^{-3}\!\!$&$\!\!\!8\!\times\! 10^{-3}\!\!$
  &$\!8\!\!\times\! 10^{-3}\!\!$&$\!\!0.90\!\!$
 &$\!\!0.32\!\!$&$\!\!0.30\!\!$\\\hline
  P3b&\!125.0\!&141.8&324.2&228.4&142.2&$0.44$&$\!1\!\!\times\! 10^{-3}\!\!\!
  $&$\!\!\!7\!\times\! 10^{-3}\!\!\!$ &$\!2\!\!\!\times\! 10^{-6}\!\!\!$
 &$\!\!0.98\!\!$&$\!\!0.10\!\!$&$\!\!0.13\!\!$\\\hline
\end{tabular}
\caption{Parameters for points P3a  and P3b.}
\label{tab:P3}
\end{table}

The first one (P3a) has a very compressed spectrum as a result of a
low value of $v_L$, as we have discussed before (see Fig.~\ref{fig:Ratios1}). 
For the second point (P3b), with a larger $v_L$, it has a broader spectrum, but the couplings to vector bosons are smaller (see Eq. (\ref{eq:3a})).

\section{Summary and conclusions}
\label{sec:Conclusions}

In this work we have examined the simplest realization of the linear
seesaw mechanism within the context of the \sm \SM gauge symmetry
structure. In addition to the standard scalar doublet, we employ two
scalar multiplets charged under lepton-number.
One of these is a nearly inert doublet, while the other is a singlet. 
Neutrino mass generation through spontaneous violation of lepton
number implies the existence of a Nambu-Goldstone boson.
The existence of such ``majoron'' would lead to stringent
astrophysical constraints. A consistent electroweak symmetry breaking
pattern requires a compressed mass spectrum of scalar bosons in which
the Standard Model Higgs boson can have a large invisible decay into the
invisible majorons.

The Higgs boson production rates are similar to what is expected
in two-doublet Higgs schemes. 
Indeed, we saw in Eq.~(\ref{eq:3a}) that the couplings of the CP even
Higgs bosons to the vector bosons obeys a simple sum rule, characteristic of
two-doublet Higgs boson schemes.
In contrast, no sum rule holds concerning their visible decay
branching, of course, so much so that, with current experimental precision, values of an invisible branching ratio up to $20\%$ are allowed. However, future
lepton colliders may play a decisive role here; in fact, it is
expected that $\text{BR}_{\text{inv}}$ may be measured with precision
better than 1\% level~\cite{Abada:2019zxq}, which will impose severe
constraints on these decay modes.
All in all, the model provides interesting and peculiar benchmarks
for electroweak breaking studies at collider experiments. We think these
deserve a dedicated experimental analysis that lies beyond the scope of this paper.


\vspace*{0.5cm}
\section*{Acknowledgments}

Work supported by the Spanish grants SEV-2014-0398 and FPA2017-85216-P
(AEI/FEDER, UE), PROMETEO/2018/165 (Generalitat Valenciana) and the
Spanish Red Consolider MultiDark FPA2017-90566-REDC. . D. F. and J. C. R are supported by projects
CFTP-FCT Unit 777 (UID/FIS/00777/2013 and UID/FIS/00777/2019), and PTDC/FIS-PAR/ 29436/2017 which
are partially funded through POCTI (FEDER), COMPETE, QREN and EU. D.F. is also supported by the Portuguese \textit{Funda\c{c}\~ao para a Ci\^encia e Tecnologia} under the project SFRH/BD/135698/2018.

\appendix

\section{Unitarity Constraints}
\label{sec:unitarity}

In this appendix we list all the coupled channel matrices for the
states in Table~\ref{tab:unitarity1}. As we discussed before, these
matrices can be separated by charge $Q$ and hypercharge $Y$ of the
initial and final state, as the states with different values of $Q,Y$
will not mix. 

\subsection{$Q=2,Y=2$}

We start with the highest charge combination. We have already
discussed this case.  We get the following matrix
\begin{equation}
  \label{eq:36}
16\pi\,  a_0^{++}=
  \begin{bmatrix}
    2 \lambda&0&0\\
    0&\beta_1+\beta_5&0\\
    0&0&2 \lambda_L
  \end{bmatrix},
\end{equation}
with eigenvalues
\begin{equation}
  \label{eq:37}
  2 \lambda, \beta_1+\beta_5, 2 \lambda_L.
\end{equation}

\subsection{$Q=1,Y=2$}

We obtain the following matrix
\begin{equation}
  \label{eq:40b}
  16\pi\, a_0^+(Y=2)=
  \begin{bmatrix}
    2\lambda &0&0&0\\
    0&\beta_1&\beta_5&0\\
    0&\beta_5&\beta_1&0\\
    0&0&0&2\lambda_L
  \end{bmatrix},
\end{equation}
with eigenvalues
\begin{equation}
  \label{eq:39b}
  2\lambda,2\lambda_L,\beta_1+\beta_5,\beta_1-\beta_5.
\end{equation}

\subsection{$Q=1,Y=1$}

We obtain the following matrix
\begin{equation}
  \label{eq:38}
  16\pi\, a_0^{+}(Y=1)=
  \begin{bmatrix}
    \beta_2 &0&0&-2\beta_4\\
    0&\beta_2&0&0\\
    0&0&\beta_3&0\\
    -2\beta_4&0&0& \beta_3
  \end{bmatrix},
\end{equation}
with eigenvalues
\begin{equation}
  \label{eq:39}
  \beta_2,\beta_3,\frac{1}{2}\left(\beta_2 + \beta_3 +
    \sqrt{16\beta_4^2+(\beta_2-\beta_3)^2}\right),
  \frac{1}{2} \left(\beta_2 + \beta_3 -
  \sqrt{16\beta_4^2+(\beta_2-\beta_3)^2}\right).
\end{equation}

\subsection{$Q=1,Y=0$}

We obtain the following matrix
\begin{equation}
  \label{eq:38}
  16\pi\, a_0^{+}(Y=0)=
  \begin{bmatrix}
    2\lambda &0&0&\beta_5\\
    0&\beta_1&0&0\\
    0&0&\beta_1&0\\
    \beta_5&0&0&2\lambda_L
  \end{bmatrix},
\end{equation}
with eigenvalues
\begin{equation}
  \label{eq:39}
  \beta_1,\beta_1,\lambda + \lambda_L +
  \sqrt{\beta_5^2+(\lambda-\lambda_L)^2}, \lambda + \lambda_L -
  \sqrt{\beta_5^2+(\lambda-\lambda_L)^2}.
\end{equation}

\subsection{$Q=0,Y=2$}

In this case we get 
\begin{equation}
  \label{eq:36}
16\pi\,  a_0^{0}(Y=2)=
  \begin{bmatrix}
    2 \lambda&0&0\\
    0&\beta_1+\beta_5&0\\
    0&0&2 \lambda_L
  \end{bmatrix},
\end{equation}
with eigenvalues
\begin{equation}
  \label{eq:37}
  2 \lambda, \beta_1+\beta_5, 2 \lambda_L.
\end{equation}

\subsection{$Q=0,Y=1$}

In this case we get the following matrix
\begin{equation}
  \label{eq:38b}
  16\pi\, a_0^{+}(Y=1)=
  \begin{bmatrix}
    \beta_2 &0&0&-2\beta_4\\
    0&\beta_2&0&0\\
    0&0&\beta_3&0\\
    -2\beta_4&0&0& \beta_3
  \end{bmatrix},
\end{equation}
with eigenvalues
\begin{equation}
  \label{eq:39b}
  \beta_2,\beta_3,\frac{1}{2}\left(\beta_2 + \beta_3 +
    \sqrt{16\beta_4^2+(\beta_2-\beta_3)^2}\right),
  \frac{1}{2} \left(\beta_2 + \beta_3 -
  \sqrt{16\beta_4^2+(\beta_2-\beta_3)^2}\right).
\end{equation}

\subsection{$Q=0,Y=0$}

Finally we get the last coupled channel matrix,
\begin{equation}
  \label{eq:41}
  16\pi a_0^0(Y=0)=
  \begin{bmatrix}
 4\lambda &0&0&\beta_{15} &2\lambda&0&0&\beta_1 &\beta_2 &0&0\\
0& \beta_{15} & 0&0 &0 &\beta_5 & 0& 0&0 &0 &-\sqrt{2}\beta_4\\
0& 0& \beta_{15} & 0&0 &0 &\beta_5 & 0& 0 &-\sqrt{2}\beta_4&0\\    
\beta_{15} & 0&0 &4\lambda_L& \beta_1& 0&0&2\lambda_L&\beta_3&0&0\\
2\lambda & 0&0 &\beta_1& 4\lambda& 0&0&\beta_{15}&\beta_2&0&0\\
0 & \beta_5&0 & 0& 0 & \beta_{15}&0&0&0&0&-\sqrt{2}\beta_4\\
0&0 &\beta_5 & 0& 0 &0 &\beta_{15} & 0& 0&-\sqrt{2}\beta_4 &0\\
\beta_1& 0& 0& 2\lambda_L& \beta_{15} &0 &0 &4\lambda_L &\beta_3 &0 &0\\
\beta_2& 0& 0&\beta_3 &\beta_2  &0 &0 &\beta_3 & 4\lambda_\sigma& 0&0\\
0&0 &-\sqrt{2}\beta_4  &0 & 0 &0 &-\sqrt{2}\beta_4  &0 &0 &2\lambda_\sigma &0\\
0&-\sqrt{2}\beta_4  &0 &0 &0  & -\sqrt{2}\beta_4 &0 &0 &0 &0 &2\lambda_\sigma
\end{bmatrix},
\end{equation}
where we have defined $\beta_{15}=\beta_1+\beta_5$. The different
eigenvalues are
\begin{align}
  \label{eq:39}
&\beta_1, \lambda+\lambda_L \pm \sqrt{\beta_5^2
  +(\lambda-\lambda_L)^2},\nonumber\\
&\frac{1}{2}\left(\beta_1 + 2 \beta_5 + 2 \lambda_\sigma \pm
  \sqrt{\beta_1^2+4 \beta_1 \beta_5-4 \beta_1 \lambda_\sigma
    +16 \beta_4^2+4 \beta_5^2-8 \beta_5
  \lambda_\sigma+4 \lambda_\sigma^2}\right) ,
\end{align}
plus the cubic roots of the equation
\begin{equation}
  \label{eq:42}
  z^3+ a z^2+ b z +c =0,
\end{equation}
where
\begin{align}
  \label{eq:43} a=&-6 \lambda-6 \lambda_{L}-4 \lambda_{\sigma},\nonumber\\
  b=&-4 \beta_{1}^2-4 \beta_{1} 
  \beta_{5}-2 \beta_{2}^2-2 \beta_{3}^2-\beta_{5}^2+36 \lambda
  \lambda_{L}+24 \lambda 
  \lambda_{\sigma}+24 
  \lambda_{L} \lambda_{\sigma} ,
  \\ c=&16 \beta_{1}^2 \lambda_{\sigma} - 8 \beta_{1} \beta_{2} \beta_{3} +
  16 \beta_{1} \beta_{5} \lambda_{\sigma} +  
 12 \beta_{2}^2 \lambda_{L} - 4 \beta_{2} \beta_{3} \beta_{5} + 12
 \beta_{3}^2 \lambda + 4 \beta_{5}^2 
 \lambda_{\sigma} -  
 144 \lambda \lambda_{L} \lambda_{\sigma}\nonumber.
\end{align}
We have implemented all the constraints that the eigenvalues should satisfy~\cite{Bento:2017eti}:
\begin{equation}
  \label{eq:44}
  |\Lambda_i| < 8 \pi \ .
\end{equation}

\bibliographystyle{JHEP}
\providecommand{\href}[2]{#2}\begingroup\raggedright\endgroup

\end{document}